\documentclass[12pt,a4paper]{article}

\newcommand{\lap}{\int_0^\infty dt e^{-st}}
\newcommand{\dmuo}{\partial^\mu}
\newcommand{\dmuu}{\partial_\mu}
\newcommand{\pkt}{\; .}
\newcommand{\kma}{\; ,}

\newcommand{\ko}{\; ,}

\newcommand{\dnuu}{\partial_\nu}
\newcommand{\intn}[2]{\int\!\frac{{ d}^{#1}\!#2}{(2\pi)^{#1}}}
\newcommand{\inte}{\int\!\frac{{ d}^{3}\!p}{ (2\pi)^{3}2E_0}}
\newcommand{\il}[2]{\int_{0}^{t#1}\!{ d}{t#2}\,}

\newcommand{\dslash}{\mbox{$\partial$\hspace{-0.55em}/}}
\newcommand{\re}[1]{{\rm Re }\,#1}
\newcommand{\im}[1]{{\rm Im }\,#1}
\newcommand{\be}{\begin{equation}}
\newcommand{\ee}{\end{equation}}
\newcommand{\bea}{\begin{eqnarray}}
\newcommand{\eea}{\end{eqnarray}}
\newcommand{\bfx}{{\bf x}}
\newcommand{\bfp}{{\bf p}}
\newcommand{\calf}{{\cal F}}
\newcommand{\caln}{{\cal N}}
\newcommand{\cale}{{\cal E}}
\newcommand{\calp}{{\cal P}}
\newcommand{\calh}{{\cal H}}
\newcommand{\cali}{{\cal I}}

\begin{document}

\begin{titlepage}
\begin{flushright}
DO-TH-98/10\\
May 1998
\end{flushright}
\vspace{20mm}
\begin{center}
{\Large \bf
Nonequilibrium dynamics of fermions in a spatially
homogeneous scalar background field}

\vspace{10mm}
{\large  J\"urgen Baacke\footnote{
 e-mail:~baacke@physik.uni-dortmund.de}, Katrin Heitmann
\footnote{e-mail:~heitmann@hal1.physik.uni-dortmund.de}, and
Carsten P\"atzold
\footnote{e-mail:~paetzold@hal1.physik.uni-dortmund.de}} \\
\vspace{10mm}

{\large Institut f\"ur Physik, Universit\"at Dortmund} \\
{\large D - 44221 Dortmund , Germany}
\vspace{15mm}

{\bf Abstract}
\end{center} 
We consider the time evolution of systems in which a 
spatially homogeneous scalar field is coupled to fermions.
The quantum back-reaction is taken into account in 
one-loop approximation. We set up the basic equations and their
renormalization in a form suitable for numerical computations.
The initial singularities appearing in the renormalized
equations are removed by a Bogoliubov transformation. 
The equations are then generalized to those in a
spatially flat Friedmann-Robertson-Walker universe.
 We have implemented the Minkowski space equations 
numerically and present 
results for the time evolution with various parameter sets.
We find that fermion fluctuations are not in general 
as ineffective as previously assumed, but show interesting
features which should be studied further. In an especially
interesting example we find that fermionic fluctuations
can ``catalyze'' the evolution of bosonic fluctuations.
\end{titlepage}

\section{Introduction}
In recent years the study of nonequilibrium dynamics in quantum field
theory has received much attention. 
Quantum fields out of equilibrium can play an essential r\^ole,
e.g., in cosmology [1-16], in the quark-gluon plasma [17-25] or during 
phase transitions in solid state physics [26].

While the formalism of nonequilibrium dynamics in quantum
field theory has been established long ago
\cite{Schwinger:1961,Keldysh:1964}, real time 
simulations for realistic systems have been developed 
only recently. 
Numerical simulations of the evolution equations have been studied
by various authors. The general features are similar:
The quantum back-reaction cannot be described
by Markovian friction terms.
The relaxation of the classical field amplitude either 
shuts off or is powerlike, the quantum ensembles generated
are characterized by parametric resonance bands, the 
full development of the resonance being suppressed by the quantum
back-reaction. The early- and late-time behavior has been
analyzed analytically for the one-loop and large-N approximations
\cite{deVega:1997,Boyanovsky:1997c}.

Most of the numerical simulations have been performed with a
scalar classical field coupled to scalar quantum fields.
The quantum back-reaction of fermion fields on a classical
scalar field has received little attention up to now.
In \cite{Boyanovsky:1995b} it has been stated, on the basis
of some numerical evidence, that at least for large
field amplitudes Pauli-blocking would make
fermions ineffective for dissipation and damping of
the classical field. In another recent publication 
\cite{Ramsey:1998} the leading orders in perturbation theory
have been evaluated, so far without numerical computations.
The interaction between a classical electric field and
fermionic fluctuations has been considered in \cite{Kluger:1992}
(see also \cite{Cooper:1994})
as a model for $q\bar q$ production in the quark-gluon plasma.
There the evolution of the system in  quantum field theory was
compared to the evolution using the Boltzmann-Vlasov equation.  
  
In this paper we reconsider fermionic fluctuations
coupled to a scalar field, with respect to both  formal
aspects and  numerical simulations.
We use a formalism developed by us recently 
\cite{Baacke:1997a,Baacke:1998b} to formulate the
renormalized equations of motion in a form which is suitable
for numerical computation but satisfies at the same time
the usual requirements of renormalized quantum field theory.
The renormalization scheme is covariant and independent
of the initial conditions.
After renormalization we find the equations to be singular
at $t=0$, a phenomenon known as Stueckelberg singularities
\cite{Stueckelberg:1951,Bogoliubov:1980}.
As in the scalar case we have studied previously
\cite{Baacke:1998a}, these
singularities can be removed by a Bogoliubov transformation
of the initial fermionic quantum state.
We then generalize the equations to those in a 
flat FRW universe. We finally formulate the linearized
equations of motion, in order to be able to compare with the
full quantum evolution. We have implemented numerically the formalism
developed in this article.
Numerical results for various parameter sets, and
various aspects of these results will be presented  and
discussed in sections 8 and in the conclusions.

The plan of this article is as follows: In section 2 we 
formulate the basic relations and the equation of motion
for a scalar field coupled to fermions; in section 3 we
present the energy momentum tensor and discuss the fermion number;
the renormalization of the equation of motion and of the
energy momentum tensor is developed in section 4; 
in section 5 we derive the Bogoliubov transformation
which removes the initial singularities; the extension to
a conformally flat Friedmann-Robertson-Walker is derived
in section 6; in section 7 we discuss the linearized
equations of motion; the results of our numerical simulations 
are discussed in section 8; conclusions are given in section 9.


\section{Basic Relations and equation of motion}
\setcounter{equation}{0}
We study a model consisting of a scalar `` inflaton'' field $\Phi$ 
coupled to a spin $1/2$ field 
$\psi$ by a Yukawa interaction. We do not introduce a genuine
mass term for the fermion field, it acquires a
time-dependent mass via the Yukawa coupling.
We introduce a $\lambda\Phi^4$ self-interaction, but do not
consider the case of spontaneous symmetry breaking and do not
consider the quantum fluctuations of the
scalar field itself. This framework
is sufficiently general for discussing renormalization and the
typical effects introduced by the back-reaction of the fermion 
fields. It can be easily generalized to include a fermion mass,
bosonic fluctuations, and spontaneous symmetry breaking. 
The Lagrangian density is given by
\begin{equation}
{\cal L}=\frac{1}{2}\dmuu\Phi\dmuo\Phi
-\frac{1}{2}M^2\Phi^2-
\frac{\lambda}{4!}\Phi^4+\overline{\psi}(i\dslash-g\Phi)\psi\kma
\end{equation}
where $M$ is the mass of the scalar field, 
and $g$ is the Yukawa coupling. 

We split the field
$\Phi$ into its expectation value $\phi$ and the quantum fluctuations
$\eta$:
\begin{equation}
\label{erw}
\Phi(\bfx,t)=\phi(t)+\eta(\bfx,t)\kma
\end{equation}
with
\begin{equation}
\phi(t)=\langle\Phi(\bfx,t)\rangle=
\frac{{\rm Tr}{\Phi\rho(t)}}{{\rm Tr}\rho(t)}\pkt
\end{equation}

The scalar fluctuations have already been 
analyzed in \cite{Baacke:1997a,Baacke:1998b}. 
The equations for the system we consider here, with the back-reaction 
of the fermion field, have been derived by \cite{Boyanovsky:1995b}
using the Schwinger-Keldysh formalism \cite{Schwinger:1961,Keldysh:1964}
and the tadpole method \cite{Weinberg:1974}. We
do not repeat it here. The equation of motion for the classical field
are given by
\be \label{phidgl}
\ddot{\phi}(t)+M^2\phi(t)+\frac{\lambda}{6}\phi^3(t)+
\frac{\lambda}{2}\langle\eta^2\rangle
+g\langle\overline{\psi}\psi\rangle=0\pkt
\ee
Here $\langle\overline{\psi}\psi\rangle$
and $\langle\eta^2\rangle$ are the expectation values of 
the quantum fluctuations of  the
fermions and the scalar field, respectively. They are related to 
CTP (closed-time-path) Green functions. They can be expressed 
by mode functions which satisfy the linearized equations of motion 
in the background field and initial conditions at some time
$t_0$. In the following we choose $t_0=0$. The scalar back-reaction
via $\langle\eta^2\rangle$ has been calculated previously
by various groups within different approximation schemes, among them large N,
Hartree, or the one-loop approximation. 
As we have explained above, here we are merely interested in the 
fermionic back-reaction and do not include the scalar one,
except for some of the numerical examples in section 8. 

The fermion field $\psi$ satisfies the Dirac equation
\be
\left[i\partial_t -{\cal H}(t)\right]\psi(t,{\bf x})=0\kma
\ee
where the Hamiltonian ${\cal H}$ is given by
\be
{\cal H}(t)=-i\mbox{\boldmath$\alpha\nabla$\unboldmath} + m(t)\beta
\pkt \ee
The term $m(t)=g\phi(t)$ is the time dependent fermion mass.
We expand the fermion field in terms of the  spinor solutions of the 
Dirac equation
\be
\psi(t,{\bf x})=\sum_s\inte\left[
b_{\bfp,s}U_{\bfp,s}(t)+d^\dagger_{-\bfp,s}{V}_{-\bfp,s}(t)
\right]e^{+i\bfp\cdot\bfx}\kma
\ee
with the time independent creation and 
annhiliation operators whose mass is determined by the initial 
state. The creation and annihilation operators
satisfy the usual anti-commutation relations
\bea
\{b_{\bfp,s},b^\dagger_{\bfp',s'}\}&=&2 E_0(2\pi)^3\delta^3(\bfp-\bfp')
\delta_{ss'}\kma\\
\{d_{\bfp,s},d^\dagger_{\bfp',s'}\}&=&2 E_0(2\pi)^3\delta^3(\bfp-\bfp')
\delta_{ss'}
\pkt
\eea
For the positive and negative energy solutions we make the ansatz
\be
U_{\bfp,s}(t)=N_0\left[i\partial_{t}+ {\cal H}_\bfp(t)\right]f_p(t)
\left(\begin{array}{c}
\chi_s \\ 0
\end{array}\right)
\ee
and
\be
V_{\bfp,s}(t)=N_0\left[i\partial_{t}+ {\cal H}_{-\bfp}(t)\right]g_p(t)
\left(\begin{array}{c}
0 \\ \chi_s
\end{array}\right)\kma
\ee
with the Fourier-transformed Hamiltonian
\be
{\cal H}_\bfp(t)=\mbox{\boldmath $\alpha p$\unboldmath} +m(t)\beta
\pkt \ee
For the two-spinors $\chi_s$ we use helicity eigenstates, i.e.,
\be
\mbox{\boldmath$ \hat p\sigma$\unboldmath} \chi_\pm=\pm \chi_\pm
\pkt
\ee
The mode functions $f_p$ and $g_p$ depend only on $p=|\bfp|$;
they obey the second order differential equations
\bea\label{fsec}
\left[
\frac{d^2}{dt^2}-i\dot{m}(t)+ p^2+m^2(t)
\right]f_p(t)&=&0\kma \\\label{gsec}
\left[
\frac{d^2}{dt^2}+i\dot{m}(t)+p^2+m^2(t)
\right]g_p(t)&=&0\pkt
\eea
The initial state for the fermion field is usually specified as
a vacuum or thermal equilibrium state 
obtained by fixing the classical field
$\phi$, and thereby the fermion mass, to some value $\phi_0$ for  $t\le 0$.
The spinor solutions 
are then identical with the usual free 
field solutions of the Dirac equation with constant mass
$m_0=m(0)=g\phi(0)$.
Therefore, the mode functions, which would be plane waves
for $t\le0$, satisfy the initial conditions
\bea\label{fginit}
f_p(0)=1&,& \dot{f}_p(0)=-iE_0\kma\\
g_p(0)=1&,& \dot{g}_p(0)=iE_0\pkt
\eea
For the spinors $U$ and $V$ we
use the usual free field  normalization conditions
\bea \label{norms}
\overline{U}_{\bfp,s}(0)U_{\bfp,s}(0)&=&
-\overline{V}_{\bfp,s}(0)V_{\bfp,s}(0)=2m_0 \kma\\
{U}^\dagger_{\bfp,s}(0)U_{\bfp,s}(0)&
=&{V}^\dagger_{\bfp,s}(0)V_{\bfp,s}(0)=2E_0\; ;
\eea
we will also need the orthogonality relation
\be \label{ortho}
{U}^\dagger_{\bfp,s}(0)V_{-\bfp,s}(0)=0
\pkt \ee
$E_0(p)$ denotes the  mode energy in the initial state, 
\be
E_0^2={{\bf p}\,}^2+m^2_0 \pkt
\ee
For the normalization constant we find
\be
N_0=\left[E_0+m_0\right]^{-1/2}\pkt
\ee
For $t > 0$ 
Eqs. (\ref{fsec}), (\ref{gsec}), and (\ref{fginit}) imply that
\be
f_p(t)=g_p^*(t)\pkt
\ee
Since the time evolution of the spinors
$U_{\bfp,s}(t)$ and $V_{-\bfp,s}(t)$ is induced by the hermitean operator
${\cal H}_\bfp$, their normalization and orthogonality
relations (\ref{norms}) and (\ref{ortho}) are conserved. 
This implies a useful relation for the mode functions
\cite{Boyanovsky:1995b}
\be \label{wronski}
|\dot f_p(t)|^2-im(t)\left[f_p(t)\dot f_p^*(t)-\dot f_p(t) f_p^*(t)\right]+
\left[p^2+m^2(t)\right]|f_p(t)|^2=2E_0\left(E_0+m_0\right)\kma
\ee
which takes the r\^{o}le of the Wronskian.
Using these mode functions $\langle\overline{\psi}\psi\rangle$ 
can be calculated once the initial state is specified
\footnote{ We use the
Heisenberg picture, i.e., the field operators depend
on time via the mode functions.}. If we use the
Fock space vacuum defined by $b_{\bfp,s}|0\rangle=0$ 
and $d_{\bfp,s}|0\rangle=0$ we get
\bea \label{f_int}
\langle
\overline{\psi}\psi\rangle
&=&\sum_s\inte\overline{V}_{-\bfp,s}(t)V_{-\bfp,s}(t)\nonumber\\
&=&-2\inte\left\{
2E_0-\frac{2\bfp^2}{E_0+m_0 }|f_p|^2
\right\}\pkt
\eea
If we use a thermal density matrix defined in terms of the
Fock space states one obtains
\be
\langle
\overline{\psi}\psi\rangle
=-2\inte \tanh\left(\frac{E_0}{2T}\right) 
\left\{2E_0-\frac{2\bfp^2}{E_0+m_0 }|f_p|^2
\right\}\; ;
\ee	
the integration measure in the momentum integrals is modified 
accordingly in the expressions for the energy-momentum tensor.

We will denote $\langle\bar\psi \psi\rangle$ as the
fluctuation integral
\be
\calf(t)=\langle\bar\psi(t) \psi(t)\rangle
\pkt \ee
The fluctuation integral is divergent and has to be regularized
and renormalized. This will be done in section 4.

\section{Energy momentum tensor and particle number}
\setcounter{equation}{0}
The energy momentum tensor of a Dirac field with mass
$m=m(t)$ is given by
\be
T_{\mu\nu}=\bar \psi \left(\frac{1}{2} i\gamma_\mu \stackrel
{\leftrightarrow}{\dnuu} +
m g_{\mu\nu}\right)\psi
\pkt\ee
The expectation value of the energy momentum tensor taken with 
the initial density matrix is spatially homogeneous and has therefore
the form $T_{\mu\nu}= {\rm diag}({\cal E},{\cal P})$. In addition to the
contribution of the Dirac field it contains the classical contribution
of the scalar field.
The energy density ${\cal E}$ of the quantum fluctuations 
is then obtained as 
\bea \label{e_int}
{\cal E}_{\rm fl}(t)&=&\langle
\overline{\psi}\left(\beta {\cal H}_p\right)\psi\rangle
=\sum_s\inte\overline{V}_{-\bfp,s}(t)
\left(\beta{\cal H}_p\right) V_{-\bfp,s}(t)
\\\nonumber
&=&2\inte\left\{i\left[E_0-m_0)\right]
\left(f_p\dot{f}^*_p-\dot{f}_pf^*_p\right)-2E_0m(t)\right\}
\pkt\eea
Using the equations of motion it is easy to see that
the time derivative of the total energy density
\be
{\cal E}={\cal E}_{\rm cl}+{\cal E}_{\rm fl}=
\frac{1}{2}\dot \phi^2(t)+\frac{1}{2}M^2\phi^2(t)
+\frac{\lambda}{4!}\phi^4(t)+{\cal E}_{\rm fl}(t)
\ee
 vanishes.
The fluctuation pressure is given by
\bea \label{p_int}
{\cal P}_{\rm fl}(t)&=&\frac 1 3\langle
\overline{\psi}\mbox{\boldmath $\gamma p$\unboldmath} \psi\rangle
=\frac 1 3\sum_s\inte\overline{V}_{-\bfp,s}(t)
\mbox{\boldmath $\gamma p$\unboldmath} V_{-\bfp,s}(t)
\\\nonumber
&=&\frac 2 3 \inte\left\{
\left[E_0-m_0\right]
\left[i\left(f_p\dot{f}^*_p-\dot{f}_pf^*_p\right)-2m(t)|f_p|^2\right]
\right\}
\kma\eea
the total pressure is
\be
{\cal P}(t)=\dot \phi^2(t)-{\cal E} +{\cal P}_{\rm fl}
\pkt\ee
Energy density and pressure are quartically divergent, their 
renormalization will be discussed in  section 4 along with
the renormalization of the fluctuation integral.

In contrast to the fluctuation integral and the energy-momentum
tensor, the definition of the particle number relies on the
creation and annihilation operators. The number of particles with momentum 
$\bfp$ and helicity $s$ is given generally
via
\be
{\cal N}_{\bfp,s}(t)\propto\langle b^\dagger_{\bfp,s}(t)b_{\bfp,s}(t)\rangle
\pkt\ee
The definition of time-dependent creation and annihilation 
operators implies an interpretation
\footnote{For an extensive discussion, in the context of
general relativity, see \cite{Birrell:1982}.}. If we used
the operators $b_{\bfp,s}$ and $b^\dagger_{\bfp,s}$ the particle number
would remain equal to the initial one, zero for an initial vacuum state,
or $1/\left[\exp(E_0/T)+1\right]$ for a thermal one. While these operators
refer to a decomposition of the field with respect to the
exact mode functions $f_p(t)$, the concept of free
particles implies plane wave mode functions. If we define
these modes to have mass $m_0$ and $E_0=\sqrt{p^2+m_0^2}$, i.e., if
we use the modes corresponding to the initial state, we obtain
\bea
b^0_{\bfp,s}(t)&=&
\int d^3xe^{-i\bfp \cdot\bfx+i E_0 t}
U^{0\dagger}_{\bfp,s}\psi(\bfx,t) \nonumber \\
&=&C^0_{\bfp, s}(t)b_{\bfp,s}+D^0_{\bfp,s}(t)d^\dagger_{-\bfp,s}
\eea
with
\be
U^0_{\bfp,s}=\frac{1}{\sqrt{E_0+m_0}}\left[E_0+ 
{\cal H}_\bfp(t_0)\right] 
\left(\begin{array}{c}
\chi_s \\ 0
\end{array}\right)\pkt 
\ee
We need only the coefficient $D^0_{\bfp,s}$ which is given by
\be \label{D_0}
D^0_{\bfp,s} = \frac{e^{iE_0t}}{2E_0}
U^{0\dagger}_{\bfp,s}V_{-\bfp,s}(t)
\kma
\ee
and the particle number becomes
\bea
{\cal N}^0 _{\bfp,s}(t)&=&|D^0_{\bfp,s}(t)|^2
\pkt
\eea
In terms of the mode functions we obtain for the occupation number
for one helicity eigenstate with momentum $\bfp$ 
\be
\caln_{\bfp}^0(t)=\frac{E_0-m_0}{4E_0^2}\left[2E_0+i[\dot f_p^*(t)f_p(t)
-f_p^*(t)\dot f_p(t)]
-2(m_1-m_0)|f_p(t)|^2
\right] \pkt
\ee
This definition has been used in \cite{Boyanovsky:1995b}. 
As it should be for fermions, the occupation number is 
strictly less or equal to $1$; this is obvious from Eq. (\ref{D_0}),
since $D^0_{\bfp,s}$ is the scalar product of two complex
vectors of unit norm. Integrating the occupation  number
over momentum one obtains the total particle number
density
\be
\caln_0(t)=\sum_s\int\frac{d^3p}{(2\pi)^3} \caln^0_{\bfp,s}(t)
\pkt\ee
If we imagine the time evolution being stopped at the time
$t$ it seems more natural to use free quanta of mass $m_1=m(t)$
and
\be
b^1_{\bfp,s}(t)=
\int d^3xe^{-i\bfp\cdot\bfx+i E_0 t}
U^{1\dagger}_{\bfp,s}\psi(\bfx,t)
\ee
with
\be
U^1_{\bfp,s}=\frac{1}{\sqrt{E_1+m_1}}\left[E_1+ 
{\cal H}_\bfp(t_1)\right]
\left(\begin{array}{c}
\chi_s \\ 0
\end{array}\right)\pkt 
\ee
The coefficient $D^1_{\bfp,s}$ now reads
\be \label{D_1}
D^1_{\bfp,s} = \frac{e^{iE_1t}}{\sqrt{4E_0E_1}}
U^{1\dagger}_{\bfp,s}V_{-\bfp,s}(t_1)
\kma\ee
and the occupation number for particles with mass 
$m_1=m(t)$ becomes 
\bea
{\cal N}^1_{\bfp,s}(t)&=&\frac{E_0-m_0}{4E_0E_1(E_1+m_1)}\biggl[
2E_0(E_0+m_0)\nonumber\\
&&+i(E_1+m_1)[\dot f_p^*(t)f_p(t)
-f_p^*(t)\dot f_p(t)]
\biggr]
\pkt\eea
The total particle number $\caln_1(t)$ is again obtained
by integrating $\caln_{\bfp,s}^1$ over momentum and summing over
helicities.
The particle number is divergent by power counting, the
analysis of the divergent contributions of $\caln_1(t)$
shows, however, that it is finite and does not
need counter terms (see also section 4). For $\caln_0(t)$ we find 
a linearly divergent contribution that vanishes in dimensional
regularization.

As we have mentioned above, the definition
of the particle number relies on an interpretation, which
seems to be more straightforward if the particle number is
computed for particles with the ``final'' mass $m_1=m(t)$.
Of course, even if the classical scalar 
field relaxes to $0$ as
$t \to \infty$, the final state (taken in the Schr\"odinger picture)
never becomes an ensemble of free particles. Such a state would
be described by a density matrix which cannot arise in 
unitary evolution from a
pure state $|0\rangle$. This is also true for the additional particles
created in the case of a thermal initial state.

Eq. (\ref{wronski}) may be used to recast the expressions for
energy density, pressure and particle number into a different,
sometimes advantageous form.


\section{Renormalization}
\setcounter{equation}{0}
In order to develop the framework for
renormalizing the one-loop equations,
we write the equation of motion for the mode 
functions, Eq. (\ref{fsec}), in the form
\be  \label{f_diff_eq}
\left[ \frac{d^2}{ dt^2}+
E_0^2\right]f_p(t)=-V(t)f_p(t)\kma
\ee
with
\be \label{pot_def}
V(t)=m^2(t)-m_0^2-i\dot{m}(t)\pkt
\ee
Using the initial conditions (\ref{fginit})
this equation can be recast into the form of
an integral equation:
\be \label{f_int_eq}
f_p(t)=e^{-iE_0 t}
-\frac{1}{E_0}\int_0^t dt' \sin[E_0(t-t')]V(t')f_p(t')
\pkt
\ee
Using this integral equation, the mode functions may be expanded
with respect to the potential $V(t)$.
We split off the zeroth order (plane wave) contribution and
an oscillating phase factor by writing
\be \label{h_def}
f_p(t)=e^{-i E_0 t}[1+h_p(t)] \pkt
\ee
The functions $h_p(t)$ satisfy differential and
integral equations derived from Eqs. (\ref{f_diff_eq})
and (\ref{f_int_eq}), respectively. These functions are
discussed in Appendix A. They may be decomposed as
\be
h_p(t)=\sum_{n=1}^\infty h_p^{(n)},
\ee
where $h_p^{(n)}$ is of $n$'th order in $V(t)$; 
we define further the inclusive sums
\be
h_p^{\overline{(n)}}=\sum_{m=n}^\infty h_p^{(m)}
\pkt \ee
In terms of these functions and their expansion discussed
in Appendix A the integrand of the fluctuation integral can be 
written as
\bea \nonumber
1-\left(1-\frac{m_0}{E_0}\right)|f_p(t)|^2&=&\frac{E_0}{m_0}
-\left(1-\frac{E_0}{m_0}\right)\left[ 2 \re h_p(t)+ |h_p(t)|^2\right]
\\ \nonumber 
&=&\frac{m(t)}{E_0}
-\frac{\ddot{m}(t)}{4(E_0)^3}-\frac{m^3(t)}{2(E_0)^3}+
\frac{m(t)m^2(0)}{2(E_0)^3}
\nonumber \\ && 
+\frac{\ddot{m}(0)}{4(E_0)^3}\cos \left(2E_0t\right)+K_{\rm F}(p,t)
\pkt
\eea
The first terms on the right hand side lead to divergent 
or singular momentum integrals. The function $K_{\rm F}(t)$ 
can be considered being defined by this
equation. It behaves as $(E_0)^{-4}$ and its momentum integral
is finite.
While $K_{\rm F}(t) $ is defined here as the difference between the 
original, numerically computed integrand
 and its leading contributions, an 
alternative expression, avoiding such a subtraction, 
is given in Appendix A.
We decompose the fluctuation integral as
\be
\calf(t)=\calf_{\rm div}(t)+\calf_{\rm sing}(t)+\calf_{\rm fin}(t)
\kma \ee
with
\bea
\calf_{\rm div}&=&\intn{3}{p}\left\{\frac{m(t)}{E_0}
-\frac{\ddot{m}(t)}{4(E_0)^3}-\frac{m^3(t)}{2(E_0)^3}+
\frac{m(t)m^2(0)}{2(E_0)^3}
\right\}\kma\\  
\calf_{\rm sing}&=&\intn{3}{p}\left\{
\frac{\ddot{m}(0)}{4(E_0)^3}\cos \left(2E_0t\right)
\right\}\kma\\
\calf_{\rm fin}&=&\intn{3}{p}K_{\rm F}(p,t)
\pkt\eea
The divergent part $\calf_{\rm div}$ is a local 
polynomial in $m(t)= g \phi(t)$. It will be absorbed by 
appropriate renormalization counter terms. The integral
involving $\cos(2 E_0 t)$ is logarithmically singular at $t=0$
but finite otherwise. This contribution is obviously related 
to the initial conditions and will be discussed in section 5.

The fluctuation parts of the energy momentum tensor can be
analyzed in a similar way.
The integrand of the energy density $\cale_{\rm fl}$ can be expanded as
\bea \nonumber
&&\frac{i}{2}\left(1-\frac{m_0}{E_0}\right)\left(
f_p\dot f_p^*-\dot f_pf_p^*\right)-m(t)\\ \nonumber
&&=-(E_0-m_0)\left \{ 1+ 2 \re h_p+|h_p|^2
+\frac{1}{E_0}\im\left[ \dot h_p\left(1+h_p^*\right)\right]\right \}
-m(t)\\ \nonumber &&=
-E_0-\frac{m^2(t)}{2E_0}+\frac{m^2_0}{2E_0}+
\frac{\dot{m}^2(t)}{8(E_0)^3}+\frac{m^4(t)}{8(E_0)^3}
 \nonumber \\ && 
\hspace{5mm}+\frac{m^4_0}{8(E_0)^3}-\frac{m^2(t)m^2_0}{4(E_0)^3}
+K_{\rm E}(p,t)\pkt
\eea
Again $K_{\rm E}(p,t)$ is defined by this equation and it behaves
as $(E_0)^{-4}$ as $E_0 \to \infty$. There is no cosine term here
and, therefore, no singular contribution.
So 
\be
\cale(t)=\cale_{\rm div}(t)+\cale_{\rm fin}(t)
\kma \ee
with
\bea
\cale_{\rm div}&=&2\intn{3}{p}
\left\{-E_0-\frac{m^2(t)}{2E_0}+\frac{m^2_0}{2E_0}+
\frac{\dot{m}^2(t)}{8(E_0)^3}+\frac{m^4(t)}{8(E_0)^3}
\right. \nonumber \\ && \left.
+\frac{m^4_0}{8(E_0)^3}-\frac{m^2(t)m^2_0}{4(E_0)^3} \right\}
\kma\\ 
\cale_{\rm fin}&=&\intn{3}{p}K_{\rm E}(p,t)
\pkt\eea
If the integrand of the fluctuation pressure is rewritten in
terms of the functions $h_p$ it reads
\be 
-(E_0-m_0)\left \{\left(1+\frac{m(t)}{4E_0}\right)
\left( 1+ 2 \re h_p+|h_p|^2\right)
-\frac{1}{E_0}\im\left[ \dot h_p\left(1+h_p^*\right)\right]\right \}
-m(t)\pkt
\ee
We expand again in $V(t)$ in order to sort out the leading
contributions.
Finally, $\calp(t)$ can be decomposed as
\be
\calp(t)=\calp_{\rm div}(t)+\calp_{\rm sing}(t)+\calp_{\rm fin}(t)
\kma \ee
with
\bea
\calp_{\rm div}&=&\intn{3}{p}\left\{
-\frac 4 3 E_0+\frac{2m_0^2}{3E_0}+\frac{m^4_0}{6(E_0)^3}+
\frac{\dot{m}^2(t)}{6(E_0)^3}-
\frac{m(t)\ddot{m}(t)} {12(E_0)^3}-\frac{m^2(t)} {3E_0}
\right.\nonumber\\
&&\left.
-\frac{m^2(t)m^2_0}{6(E_0)^3}
\right\}\kma
\\
\calp_{\rm sing}&=&\intn{3}{p}
\frac{m(t)\ddot{m}(0)}{12(E_0)^3}\cos(2E_0t)
\kma\\
\calp_{\rm fin}&=&\intn{3}{p} K_{\rm P}(p,t)
\pkt\eea
The integral over $K_{\rm P}$, which is defined by this decomposition,
is finite.
The divergent terms $\calf_{\rm div},\cale_{\rm div}$,
 and $\calp_{\rm div}$ are proportional to local terms in
$\phi(t)$ and its derivatives. These can be absorbed in the usual
way by introducing the appropriate counter terms into the Lagrangian
and into the energy-momentum tensor.

The counter terms in the Lagrangian are introduced as
\be
{\cal L}_{\rm c.t}=\frac 1 2\delta Z\dot{\phi}^2-
\frac 1 2 \delta M^2\phi^2-\frac{\delta
\lambda}{24}\phi^4\pkt
\ee
The divergent parts of the fluctuation integral
can be evaluated, e.g., using dimensional regularization.
One finds
\be
\calf_{\rm div}=\frac{\ddot{m}(t)}{8\pi^2}L_0+\frac{m^3(t)}{4\pi^2}L_0
+\frac{m(t)m_0^2}{4\pi^2}\kma
\ee
with the abbreviation
\be
L_0= \frac{2}{\epsilon}+
\ln{\frac{4\pi\mu^2}{m_0^2}}-\gamma\pkt
\ee  
As already found for the scalar fluctuations
\cite{Baacke:1997a}, the dependence 
on the initial mass $m_0$ can be absorbed into finite terms,
$\Delta Z,\Delta M^2$ and $\Delta \lambda$. 
Applying an $\overline{MS}$ prescription,
the infinite renormalizations become
\bea
\delta Z&=&-\frac{g^2}{8\pi^2}L\kma\\
\delta \lambda&=&-6\frac{g^4}{4\pi^2}L
\kma \eea
with 
\be
L= \frac{2}{\epsilon}+
\ln{\frac{4\pi\mu^2}{M^2}}-\gamma\pkt
\ee
There is no infinite mass renormalization counter term.
Introducing the renormalization counter terms into the equation
of motion, we obtain
\be \label{ren_eqm}
\left(1+\Delta Z\right)\ddot{\phi}+ \left(M^2+
\Delta M^2\right)\phi
+\frac{\lambda+
\Delta \lambda}{6}\phi^3+g(\calf_{\rm fin}
+\calf_{\rm sing})=0
\pkt
\ee
The coefficients of the finite terms left over after adding the
renormalization counter terms to $\calf_{\rm div}$
are given by
\bea
\Delta Z&=&\frac{g^2}{8\pi^2}\ln\frac{M^2}{m_0^2} \kma \\
\Delta \lambda&=&6\frac{g^2}{4\pi^2}\ln\frac{M^2}{m_0^2}\kma \\
\Delta M^2&=&\frac{g^2 m_0^2}{4\pi^2}\pkt
\eea
Since the bare fermion mass vanishes, we have introduced 
the scalar mass $M$ as scale parameter.
Obviously, the equation of motion is not yet acceptable in its present 
form, due to the singular term.

The divergent parts of the energy give, after dimensional regularization
\be
{\cal E}_{\rm div}=\frac{\dot{m}^2(t)}{16\pi^2}L_0+
\frac{m^4(t)}{16\pi^2}L_0-\frac{m^4(0)}{32\pi^2}
+\frac{m^2(t)m^2_0}{8\pi^2}\pkt
\ee
The counter terms correspond to those in the Lagrangian, 
i.e.,
\be
{\cal E}_{\rm c.t}=\frac 1 2\delta Z\dot{\phi}^2+
\frac 1 2 \delta M^2\phi^2+\frac{\delta
\lambda}{24}\phi^4 
\ee
with the same coefficients as above.
We need no infinite counter term for the zero point energy
or cosmological constant.
Adding the divergent part and the counter terms we are 
left with finite contributions
\be
\cale_{\rm div}+\cale_{\rm c.t.}=
\frac{1}{2}\Delta Z \dot \phi^2 +\frac{1}{2}\Delta M^2
\phi^2 + \frac{\Delta \lambda}{24}\phi^4+\Delta \Lambda
\kma \ee
with
\be
\Delta\Lambda=-\frac{m_0^4}{32 \pi ^2}
\pkt
\ee
The divergent part of the pressure is given by
\be
{\cal P}_{\rm div}=\frac{\dot{m}^2(t)}{12\pi^2}L_0-
\frac{m(t)\ddot{m}(t)}{24\pi^2}L_0-\frac{m_0^4}{24 \pi^2}
+\frac{m_0^2m^2(t)}{12\pi^2}
\pkt\ee
In addition to the counter terms already introduced we have to
add to the energy momentum tensor the ``improvement'' counter term
\cite{Callan:1970}
\be
\delta A(g_{\mu\nu}\partial^\alpha\partial_\alpha-
\dmuu\dnuu)\phi^2(x)\; ;
\ee
since $\phi$ depends only on $t$ this term
contributes only to the pressure, and we have
\be
{\cal P}_{\rm c.t}=
+\delta Z\dot{\phi}^2+
\delta A\frac{d^2}{dt^2}\phi^2\pkt
\ee
We choose
\be
\delta A=\frac{g^2}{48\pi^2}L \; ;
\ee
there is a finite remainder
\be
\calp_{\rm div}+\calp_{\rm c.t.}=
+\Delta Z\dot{\phi}^2+
\Delta A\frac{d^2}{dt^2}\phi^2
-\frac{m_0^4}{24 \pi^2}
+\frac{m_0^2m^2(t)}{12\pi^2}
\kma \ee
with
\be
\Delta A = \frac{g^2}{48 \pi^2}\ln\frac{M^2}{m_0^2}
\pkt
\ee 


\section{Removing the initial singularity}
\setcounter{equation}{0}
We are now ready to discuss the terms proportional
to $\cos(E_0 t)$ which turn out to be singular
as $t \to 0$ :
\be
\calf_{\rm sing}(t)=
\intn{3}{p}\frac{\ddot{m}(0)}{4(E_0)^3}\cos(2 E_0 t) \simeq 
-\frac{\ddot{m}(0)}{16\pi^2}\ln (2 m_0 t) \; \
{\rm as} \;\ t \to 0 \pkt
\ee
For the case of scalar fields in Minkowski space
 the fluctuation integral
has only a logarithmic cusp $\propto t \ln t$ at $t=0$,
as observed by Ringwald \cite{Ringwald:1987}; the
energy is finite, and the pressure behaves
as $\ln t$. In FRW cosmology the energy of scalar fluctuations
is logarithmically singular whereas the pressure behaves as
$1/t$. So the Friedmann equations become singular. 
Problems with the initial conditions in FRW cosmology
 have also been noted
in \cite{Boyanovsky:1994,Boyanovsky:1997a} when using comoving
time and the associated vacuum state; the problems disappear 
if conformal time is used. The two
vacuum states are related by a Bogoliubov transformation.

For fermionic fluctuations  we find that already the fluctuation 
integral is divergent, so that 
the numerical code cannot be started even in Minkowski space.
We have shown recently,
for the case of scalar fields, that such ``Stueckelberg singularities''
\cite{Stueckelberg:1951} can be removed by a Bogoliubov 
transformation of the initial state, which was constructed
explicitly.

Within the Fock space based on the ``initial vacuum'' state
$|0\rangle$ which is annihilated by the operators
 $b_{\bfp,s}$ and $d_{\bfp,s}$ 
we define a more general initial state by requiring that
\be
\left[b_{\bfp,s}-\rho_{\bfp,s} d^\dagger_{-\bfp,s}\right]
|\tilde 0\rangle =0 \pkt
\ee
The Bogoliubov transformation from $|0\rangle$ to this state
is given in Appendix B.
If the fluctuation integral, the energy, and the pressure are
computed by taking the trace with respect to this
state, we just have to replace in the defining equations
(\ref{f_int}),(\ref{e_int}) and (\ref{p_int})
the functions $U_{\bfp,s}(t)$ by
\be
V_{-\bfp,s}(t) \Rightarrow \cos(\beta_{\bfp,s})V_{-\bfp,s}(t)+
\sin(\beta_{\bfp,s})U_{\bfp,s}(t) \pkt
\ee
For the particle number, the substitution is done in the
Bogoliubov coefficients, Eqs. (\ref{D_0}) and (\ref{D_1}).
The angle $\beta_{\bfp,s}$ is related to $\rho_{\bfp,s}$ 
via
\be
\rho_{\bfp,s}=\tan(\beta_{\bfp,s})\pkt
\ee
If the expectation value of $\bar\psi\psi$ is taken in the
Bogoliubov-rotated initial state the fluctuation integral
becomes
\bea
\tilde \calf(t)&=&\sum_s\inte\left\{
\overline{V}_{-\bfp,s}(t)V_{-\bfp,s}(t)\cos^2\beta_{\bfp,s}
+\overline{U}_{\bfp,s}(t)U_{-\bfp,s}(t)\sin^2\beta_{\bfp,s}\right.\nonumber\\
&&\left.+\left[\overline{U}_{\bfp,s}(t)V_{-\bfp,s}(t)
+\overline{V}_{\bfp,s}(t)U_{-\bfp,s}(t)\right]
\sin\beta_{\bfp,s}\cos\beta_{\bfp,s}
\right\}\pkt
\eea
Rewriting this expression in terms of the mode functions 
we find
\bea
\tilde \calf(t)&=&
-\sum_s\inte\left\{\cos 2\beta_{\bfp,s}\left[2E_0-
\frac{2{\bf{p}}^2}{E_0+m_o}|f_p|^2\right]\right.\nonumber\\
&&\left.+\sin 2\beta_{\bfp,s}\frac{sp}{E_0+m_0}
\left[\im \partial_t f_p^2-2m(t)\re f_p^2 \right]
\right\}\pkt
\eea
Using the perturbative expansion of the mode 
functions given in Appendix A this integral takes the form
\bea
\tilde\calf(t)&=
&-\sum_s\intn{3}{p}\left\{\cos 2\beta_{\bfp,s}\left[
\frac{m(t)}{E_0}
-\frac{\ddot{m}(t)}{4(E_0)^3}-
\frac{m^3(t)}{2(E_0)^3}+\frac{m(t)m^2(0)}{2(E_0)^3}
\right.\right.\nonumber\\
&&\left.
+\frac{\ddot{m}(0)}{4(E_0)^3}\cos \left(2E_0t\right)+
K_{\rm F}(p,t)
\right]\nonumber\\\label{psibarpsidiv}
&&\left.+\sin 2\beta_{\bfp,s}\frac{sp}{2E_0(E_0+m_0)}\left[
-2E_0\cos 2E_0t+L_{\rm F}(p,t)\right]\right\}\pkt
\eea
The functions $K_{\rm F}(p,t)$ has been defined above,
$L_{\rm F}(p,t)$ is defined by
\be
L_{\rm F}(p,t)=
2\im\left[e^{-2iE_0t}\left((1+h_p)\dot h_p-iE_0(2h_p+h_p^2)\right)
\right]-2m(t)\re f_p^2 \pkt
\ee
Obviously, one gets rid of the term proportional to $\cos(2E_0t)$
by requiring
\be
\tan 2\beta_{\bfp,s}=
\frac{\ddot{m}(0)(E_0+m_0)}{8sp(E_0)^3}\pkt
\ee
Thereby, the Bogoliubov transformation is explicitly specified.
We notice that the helicity dependence
of the Bogoliubov transformation cancels in Eq. (\ref{psibarpsidiv}).
Using the asymptotic behavior
\be
\beta_{\bfp,s}
\stackrel{p\to\infty}{\simeq}\frac{\ddot{m}(0)}{8sp(E_0)^2}
\kma \ee
and therefore
\be\label{cosbeta} 
\cos(2\beta_{\bfp,s})-1=2\sin^2 (\beta_{\bfp,s})\stackrel{p\to\infty}{\simeq}
\frac{\ddot{m}^2(0)}{64p^2(E_0)^4}\kma
\ee
\be 
\sin(2\beta_{\bfp,s})\stackrel{p\to\infty}{\simeq}
\frac{\ddot{m}(0)}{4sp(E_0)^2}\kma
\ee
it is easy to convince oneself that this Bogoliubov transformation
does not interfere with the analysis of the divergent parts and, therefore, 
with the renormalization discussed in the previous section.
So $\tilde \calf$ is rendered finite by adding the counter
terms defined in the previous section. In the renormalized equation of motion
(\ref{ren_eqm}) we just have to replace
$\calf_{\rm sing}(t)+\calf_{\rm fin}(t)$ 
with $\tilde\calf_{\rm fin}(t)$ 
which is given explicitly by
\bea
\tilde\calf_{\rm fin}(t) &=&-2\intn{3}{p}\left\{
-2\sin^2\beta_{\bfp,s}\left[
\frac{m(t)}{E_0}
-\frac{\ddot{m}(t)}{4(E_0)^3}-\frac{m^3(t)}{2(E_0)^3}+\frac{m(t)m^2(0)}{2(E_0)^3}
\right]\right.\nonumber\\
&&\left.+\cos 2\beta_{\bfp,s} K_{\rm F}(p,t)+
\sin 2\beta_{\bfp,s}\frac{sp}{2E_0(E_0+m_0)}L_{\rm F}(p,t)
\right\}\pkt
\eea
The renormalization of the energy density proceeds in an analogous way.
The fluctuation energy in the Bogoliubov transformed state is
\bea\nonumber
\tilde {\cal E_{\rm fl}}&=&\sum_s\inte\left\{
\cos^2\beta_{\bfp.s}\,
\overline{V}_{-\bfp,s}(t)\left(\beta{\cal H}_\bfp\right)V_{-\bfp,s}(t)
\right.
\\  &&+\sin^2\beta_{\bfp,s}\,\overline{U}_{\bfp,s}(t)
\left(\beta{\cal H}_\bfp\right)U_{-\bfp,s}(t)\\\nonumber
&&\left.+\sin\beta_{\bfp,s}\cos\beta_{\bfp,s}\left[\overline{U}_{\bfp,s}(t)
\left(\beta{\cal H}_p\right)V_{-\bfp,s}(t)
+\overline{V}_{\bfp,s}(t)\left(\beta{\cal H}_p\right)U_{-\bfp,s}(t)\right]
\right\}\pkt
\eea
We again insert the expansion of the mode functions to obtain
\bea
\tilde {\cal E}_{\rm fl}&=&\sum_s\inte\left\{\cos 2\beta_{\bfp,s} 
\left[i\left[E_0-m(0)\right]
\left(f_p\dot{f}^*_p-\dot{f}_pf^*_p\right)-2E_0m(t)\right]
\right.\nonumber\\
&&\left.+\sin 2\beta_{\bfp,s} 
\frac{sp}{E_0+m_0}\re\left[\dot{f}^2_p+E^2f_p^2\right]
\right\} \pkt
\eea
After adding the counter terms defined in the previous section
the finite part of the fluctuaion energy becomes
\bea
\tilde {\cal E}_{\rm fin}&=&2\intn{3}{p}\left\{
-2\sin^2\beta_{\bfp,s}\left[
-E_0-\frac{m^2(t)}{2E_0}+\frac{m^2_0}{2E_0}+
\frac{\dot{m}^2(t)}{8(E_0)^3}
\nonumber\right.\right.\\&&\left.
+\frac{m^4(t)}{8(E_0)^3}
+\frac{m^4_0}{8(E_0)^3}-\frac{m^2(t)m^2_0}{4(E_0)^3}\right]
\\\nonumber&&\left.
+\cos 2\beta_{\bfp,s} K_{\rm E}(p,t)+ 
\sin 2\beta_{\bfp,s} \frac{sp}{2 E_0\left(E_0+m_0\right)}L_{\rm E}(p,t)
\right\}\kma
\eea
with 
\be
L_{\rm E}(p,t)=\re\left\{e^{-2iE_0t}
\left[\dot{h_p}^2-2 i E_0\dot{h_p}\left(1+h_p\right)
+ (m^2(t)-m_0^2)(1+h_p)^2\right]\right\}\pkt
\ee
Finally, we consider the fluctuation pressure, taken in the new vacuum state.
 It reads
\bea
\tilde\calp_{\rm fl}&=&-\tilde \cale_{\rm fl}+\sum_s\intn{3}{p}\left\{
\cos 2\beta_{\bfp,s}\left[
\frac 4 3 i\left(f_p\dot{f}_p^*-\dot{f}_p f_p^*\right)
\left(E_0-m_0\right)\right.\right.\nonumber\\
&&\left.-\frac 2 3 m(t)
\left(E_0-m_0\right)|f_p|^2
-2m(t)E_0\right]
\\ \nonumber&&\left.+\sin 2 \beta_{\bfp,s}\frac{sp}{E_0+m_0}
\re\left[\frac 4 3 \dot{f}_p^2-\frac 1 3 im(t)\partial_t
f_p^2+\left(\frac 4 3 {\bf{p}}^2+\frac 2 3 m^2(t)\right)f_p^2\right]
\right\}\pkt
\eea
The finite parts becomes, after adding the counter terms,
\bea
\tilde {\cal P}_{\rm fin}&=&2\intn{3}{p}\left\{-2\sin^2\beta_{\bfp,s}
\left[-\frac 4 3 E_0+\frac{2m_0^2}{3E_0}+\frac{m^4_0}{6(E_0)^3}+
\frac{\dot{m}^2(t)}{6(E_0)^3}\right.\right.\nonumber\\
&&-\left.\frac{m(t)\ddot{m}(t)} {12(E_0)^3}
-\frac{m^2(t)} {3E_0}-\frac{m^2(t)m^2_0}{6(E_0)^3}
\right]\\\nonumber
&&\left.+\cos 2 \beta_{\bfp,s} K_{\rm P}(p,t)+\sin 2 \beta_{\bfp,s}
 \frac{sp}{2E_0\left(E_0+m_0\right)}L_{\rm P}(p,t)
\right\}
\eea
with
\bea
L_{\rm P}(p,t)&=&\frac{1}{3}\re\left\{4 \dot f_p^2-2 i m 
e^{-2iE_0t}\left[\dot h_p (1+h_p)-i E_0(2h_p+h_p^2)\right]
\right.\nonumber\\&&\left.+
\left[4 \bfp^2+2 m^2(t)\right]f_p^2\right\}
\pkt\eea
The Bogoliubov transformation has removed the singular term
in the pressure as well.


\section{Extension to FRW spacetime}
\setcounter{equation}{0}
Now that we have set all basic 
equations and performed renormalization the extension to FRW spacetime
is straightforward.

We consider the Friedmann-Robertson 
Walker (FRW) metric with curvature parameter $k=0$, i.e., a 
spatially isotropic and flat spacetime.
We will treat the the quantum fields and the 
cosmological background self-consistently. That is, the
scale parameter $a(t)$ is obtained dynamically from the quantum fields.

The line-element of a flat FRW universe is given by
\be
\label{line}
{\rm d}s^2={\rm d}t^2-a^2(t){\rm d}{\vec x~}^2\pkt
\ee
The time evolution of the $a(t)$ is governed
by Einstein's field equation
\be\label{Einfield}
G_{\mu\nu}+\alpha  H^{(1)}_{\mu\nu}+\beta  H^{(2)}_{\mu\nu}
+\Lambda g_{\mu\nu}=
-\kappa
\langle T_{\mu\nu}\rangle
\ee
with $\kappa= 8\pi G$.
The Einstein curvature tensor $G_{\mu\nu}$ is given by
\be
G_{\mu\nu}=R_{\mu\nu}-\frac 1 2 g_{\mu\nu}R\pkt
\ee
The Ricci tensor and the Ricci scalar are defined as
\begin{eqnarray}
R_{\mu\nu}&=&R^{\lambda}_{\mu\nu\lambda}\kma\\
R&=&g^{\mu\nu}R_{\mu\nu}\kma
\end{eqnarray}
where
\be
R^{\lambda}_{\alpha\beta\gamma}=
\partial_{\gamma}\Gamma^{\lambda}_{\alpha\beta}
-\partial_{\alpha}\Gamma^{\lambda}_{\gamma\beta}
+\Gamma^{\lambda}_{\gamma\sigma}\Gamma^{\sigma}_{\alpha\beta}
-\Gamma^{\lambda}_{\alpha\sigma}\Gamma^{\sigma}_{\gamma\beta}\pkt
\ee
The terms  $H_{\mu\nu}^{(1)}$ and $H_{\mu\nu}^{(2)}$ arise if
terms proportional to $R^2$ and $R^{\mu\nu}R_{\mu\nu}$ are
included into the Hilbert-Einstein action.
If space-time is conformally flat, these terms are related by
\be
 H^{(2)}_{\mu\nu}=\frac{1}{3} \mbox{ }H^{(1)}_{\mu\nu}\kma
\ee
so that we can set $\beta=0$ 
in (\ref{Einfield})
without loss of generality \cite{Birrell:1982}.
We also replace $H_{\mu\nu}^{(1)}$
by $H_{\mu\nu}$ in the following.

These terms are usually not considered in standard cosmology.
They are included here, as well as
the cosmological constant term, only for the purpose of
renormalization; they will absorb divergences of the energy momentum
tensor. So in principle they should appear on the right hand side
as counterterms;
they are related to the coefficients of these
counterterms  by $\Lambda=\kappa\delta\Lambda$ and
$\alpha=\kappa\delta\alpha$.

As usual we can reduce the Einstein field equations to an equation
for the time-time component and one for
the trace of $G_{\mu\nu}$, the Friedmann
equations
\bea
\label{Friedtt}
G_{tt}+ \alpha H_{tt}+\Lambda &=&-\kappa T_{tt}\kma\\
\label{Friedtr}
G_\mu^\mu+ \alpha H^{\mu}_\mu+4\Lambda &=&-\kappa T^{\mu}_{\mu}\pkt
\eea
For the line element (\ref{line}) the various terms take the form
\cite{Birrell:1982}
\bea
G_{tt}(t)&=&-3H^2(t)\kma \nonumber\\
G_{\mu}^\mu(t)&=&-R(t)  \kma\nonumber \\
H_{tt}(t)&=&-6\left(H(t)\dot{R}(t)+H^2(t)R(t)
-\frac{1}{12} R^2(t)\right)\kma\nonumber\\
H^{\mu}_\mu(t)&=&-6\left(\ddot{R}(t)+3H(t)\dot{R}(t)\right)
\kma \eea
with the curvature  scalar
\be
R(t)=6\left[\dot{H}(t)+2H^2(t)\right]
\ee
and the Hubble expansion rate 
\be
H(t)=\frac{\dot{a}(t)}{a(t)}\pkt
\ee
The Dirac equation in FRW spacetime (see, e.g.,
\cite{Birrell:1982,Barut:1987}) is given by	
\be
\left\{i\partial_t+i\frac{3}{2}\frac{\dot{a}(t)}{a(t)}+
\frac{i}{a(t)}\mbox{\boldmath $\alpha \nabla$\unboldmath} 
-g \phi(t)\gamma_0
\right\}\psi(t,{{\bf x}})=0
\pkt \ee
It proves convenient to introduce conformal time
and scales.
The conformal factors 
for the scalar field and the fermion field are:
\bea
\psi(t,\bfx)&=&a^{-3/2}(t)\tilde\psi(\tau,\tilde\bfx)\kma \\
\phi(t,\bfx)&=&a(t)^{-1}\tilde\phi(\tau,\tilde\bfx)\kma
\eea
with $\bfx=a(t)\tilde\bfx$ and $dt=a(t)d\tau$.
In conformal time, and using these redefinitions of the
fields, the Dirac equation simplifies to
\be
\left\{i\partial_{\tau} +i
\mbox{\boldmath$\alpha$\unboldmath}\tilde{\bf \nabla} -g 
\tilde\phi(\tau)
\beta\right\}\tilde\psi(\tau ,\tilde \bfx)=0 \pkt
\ee 
If we introduce, as in section 2, the Dirac Hamiltonian 
\be
\tilde{\cal H}(\tau)=-i\mbox{\boldmath$\alpha$\unboldmath}
\tilde{\bf \nabla} +g\tilde\varphi(\tau)\beta
\pkt \ee
this equation takes the standard form
\be
\left[i\partial_\tau -\tilde{\cal H}(\tau)\right]
\tilde\psi(\tau,\tilde\bfx)=0
\pkt
\ee
So the formalism of quantization can be taken over from 
section 2,  if the other functions and operators are
understood as being rescaled quantities as well. 
The rescalings for the other quantities of interest are,
 omitting arguments, super- and
subscripts:
\bea
\begin{array}{lcl}
x = a \tilde x &,& p = a^{-1} \tilde p \kma  \\
b = a \tilde b &,& d= a \tilde d \kma \\
E_0= a^{-1}\tilde E_0 &,& m=a^{-1}\tilde m=g \tilde\phi\kma \\
U=a^{-1/2}\tilde U &,& V=a^{-1/2}\tilde V\kma
\\
\calh = a^{-1}\tilde \calh &,& N_0 = a^{1/2}\tilde N_0
\pkt\end{array}
\eea
The potential $V(t)$ , Eq. (\ref{pot_def}), 
becomes the analogous expression
$\tilde V(\tau)$ with $m(t)$ replaced by $\tilde m(\tau)$.
This means that the entire perturbative expansion and
the analysis of divergences proceeds in perfect analogy
to the Minkowski space analysis. The metric does not appear
in this formalism, and therefore there are no divergences
related to the metric. So the Dirac field does not contribute
to the wave function renormalization of the
gravitational field $\delta Z_g$ (or, equivalently, the
renormalization of Newtons constant) and to
the terms $H^{(i)}_{\mu\nu}$. 

In the equation of motion the fluctuation integral scales 
as $a^{-3}$ and so do the kinetic  and the $\lambda \phi^3$ term.
Therefore, the divergent parts of the 
fluctuation integral are
absorbed by exactly the same counter terms $\delta Z \ddot \phi$
and $\delta \lambda \phi^3/6$ as in  Minkowski 
space. The same holds true for
the finite remainders proportional to
$\Delta Z$ and $\Delta \Lambda$. However, the
finite mass renormalization $\Delta M^2 \phi$ is now replaced by
$a^{-3}\Delta M^2 \tilde \phi$ while the genuine mass term scales
as $a^{-1}M^2 \tilde \phi$.

The renormalized equation of motion for the scalar field $\tilde{\phi}$
takes therefore the form
\be
(1+\Delta Z)\frac{d^2}{d\tau^2} \tilde \phi +
a^2\left[M^2+\left(\xi-\frac 1 6\right) R\right] \tilde
\phi + \Delta M^2 \tilde \phi
 + \frac{\lambda+\Delta \lambda}{6}\tilde \phi^3 + g \tilde 
\calf_{\rm fin} =0
\pkt\ee

The energy-momentum tensor generated by the fermionic fluctuations
scales exactly as $a^{-4}$.


\section{The linearized equations of motion}
A simple intuitive approach to the interplay between the classical
Higgs field oscillating with a frequency of the order
of the Higgs mass $M$ and the fermionic fluctuations is
to treat the system as a Higgs field decaying at rest into
the fermions. For large amplitudes of the Higgs field
this picture is certainly inadequate. 
For small amplitudes, e.g., at the
end of inflation, the linearization of the equations of motion
leads indeed to an approximation which supports the simple
decay picture. We will compare the exact equations and the
lowest approximation in our numerical examples. Here we analyze
the behavior of the system analytically along the lines in 
\cite{Boyanovsky:1995b}.

If we retain only the $O(g^2)$ part of the fluctuation integral,
we find a divergent part which is cancelled by the 
wave function renormalization counter term, a finite part
remains and the linearized equation of motion of
the classical field takes the form
\be
(1+\Delta Z)\ddot \phi(t)+M^2 \phi(t) 
+\int_0^t dt'\Sigma_r(t-t')\stackrel{\dots}{\phi}(t')
+\Sigma_r(t)\ddot\phi(0) =0
\pkt\ee
The self-energy insertion is subtracted 
and is given explicitly by
\be \label{sek}
\Sigma_r(t,t')=-2g^2\int\frac{d^3p}{(2\pi)^3}\frac{1}{4 E_0^3}
\cos(2 E_0 (t-t'))
\pkt\ee
We define the Laplace transform of the condensate $\phi(t)$ via
\be
\psi(s)=\lap\phi(t)
\kma\ee
the inverse transformation being given by
\be
\phi(t)=\frac{1}{2\pi i} \int_{-i\infty+c}^{i\infty+c}
e^{st} \psi(s)
\pkt
\ee
The ``Bromwich''-contour of the latter integral runs parallel
to the imaginary axis. The constant $c$ has to be chosen
in such a way that $\psi(s)$ is analytic for $\re s > c$.
In our application $\psi(s)$ will have 
cuts along the imaginary axis and poles in the half-plane
$\re s < 0$.

For the Laplace transform $\psi(s)$ the equation of motion 
reads
\bea
&&(1+\Delta Z)\left[s^2\psi(s)-s\phi(0)-\dot\phi(0)\right]
\\ \nonumber
&&\hspace{4mm} +M^2 \psi(s) +\tilde \Sigma_r(s) 
\left[-\ddot\phi(0)-s\dot\phi(0)-s^2\phi(0)
+s^3\psi(s)\right]+\tilde\Sigma_r(s)\ddot\phi(0)=0
\kma\eea
where $\tilde \Sigma(s)$ is the Laplace transform of the self-energy
kernel. The equation can be solved readily with the result
\be \label{psi}
\psi(s)=\frac{\left[\dot\phi(0)+s\phi(0)\right]
\left[1+\Delta Z +s\tilde\Sigma_r(s)\right]}
{(1+\Delta Z) s^2 + M^2 +s^3\tilde\Sigma_r(s)}
\pkt\ee
The singularities of the right hand side in the complex $s$ plane
are given, on the one hand, by the singularities
of $\Sigma_r(s)$, and by
possible zeros of the denominator, whose locations have to 
be determined. Since $\phi(t)$ cannot contain contributions
that increase exponentially, these poles will have to lie
on the left part of the complex plane. The dominant behavior
at large $t$ will be governed, therefore, by the singularities 
of $\tilde\Sigma_r(s)$ on the imaginary axis.

The self energy kernel has been defined in Eq. (\ref{sek}).
Its Laplace transform is given by
\be
\Sigma_r(s)=-2g^2\int\frac{d^3p}{(2\pi)^3}\frac{1}{E_0^3}
\frac{s}{s^2+4E_0^2}
\pkt\ee
It has cuts along the imaginary $s$-axis running from
$s=4 i m_0$ to $s=i\infty$, and from
$s=-4 i m_0$ to $s=-i\infty$. We introduce the frequency
$\omega=-is$. Expressed in the variable $\omega$ the cuts
run from $4m_0$ to $\infty$ and from $-\infty$ to $-4m_0$. 
The discontinuity across the cut along the positive
imaginary axis is defined as
\be
\rho(s)={\rm disc}\tilde\Sigma_r(s)=\tilde\Sigma_r(i\omega+\epsilon)
-\tilde\Sigma_r(i\omega-\epsilon)
\pkt\ee
One finds
\be
\rho(s)=-2g^2\int\frac{d^3p}{(2\pi)^3}\frac{1}{4E_0^3}
\frac{2\pi \omega}{\omega+2E_0}\delta(\omega-2E_0)
=-\frac{g^2}{4\pi}\frac{\sqrt{\omega^2-4 m_0^2}}{\omega^2}
\pkt\ee
A more relevant quantity for $\psi(s)$ is the discontinuity
of the denominator of Eq. (\ref{psi}):
\be
i \gamma(s)=\left[{\rm disc}s^3\Sigma(s)\right]|_{s=i\omega}=
i \frac{g^2}{4\pi}\omega\sqrt{\omega^2-4 m_0^2}
\pkt\ee
The discontinuity of the denominator is purely imaginary, as
to be expected. If $\Sigma$ is small, we can expand the denominator
around its zero at $s \simeq iM$. The contribution of this pole
is related to the decay of the condensate particles of mass $M$ 
into fermions of mass $m_0$. We neglect the real part
of $\Sigma$ as it just shifts the value of M. Indeed it should
be zero at $s=\pm i M$ if the condensate field
is renormalized on shell. We write
\be
s^2+(M^-i\Gamma/2)^2 \simeq -(\omega^2-M^2+iM\Gamma)
\simeq s^2+M^2-i\gamma(M)/2
\pkt
\ee
It follows:
\be \label{gamma}
\Gamma=\frac{1}{2M}\gamma(M)=\frac{g^2}{8\pi}\sqrt{M^2-4 m_0^2}
\pkt\ee
This is almost, but not quite, what one would expect for fermions. Evaluating
the width of $\phi\to f\bar f$ in the standard way one finds
\be
\Gamma_{\phi\to f\bar f}=\frac{g^2}{8\pi M}\left(M^2-4 m_0^2
\right)^{3/2}
\pkt\ee
A factor $p$ is typical for an s-wave, the factor $p^3$ in the
correct decay formula arises from the Dirac traces.
While at first glance the difference is surprising, one may
recognize that the factor, by which the two expressions
differ, is given by $1-4 m_0^2/M^2$. The fermion mass is,
however, of order $g^2$, so that the difference is higher order
and our approximation to the full one-loop result is
just lowest order in $g^2$ only.

The late-time behavior of $\phi(t)$ is determined by the strongest 
singularities of its Laplace transform. These are, on the one hand,
the poles at  $s=\pm iM-\Gamma/2$ and, on the other hand,
the branch cuts on the imaginary axis.
The contribution of the poles, which are actually in the second
Riemann sheet, has been analyzed carefully in \cite{Boyanovsky:1995b}.
If the poles have residue $R$ they contribute
\be \label{exp_damp}
\phi(t)\simeq\frac{1}{2\pi i} 2 \pi i R \left(e^{iM-\Gamma/2t}
+e^{-iM-\Gamma/2t}\right)=
R e^{-\Gamma t/2}2 \cos (Mt)
\pkt\ee
Approximately, $R\simeq \phi(0)/2$.
While this contribution decreases exponentially,
the singularities on the imaginary axis
yield a power behavior
\be \label{pow_damp}
\phi(t)= R t^{-\alpha} \cos(\Omega t+\varphi)
\kma\ee
where the power-like decrease as $t^{-\alpha}$ is related to
the order of the branch point $-1+\alpha$.
Since the branch point is of the square root type
we have $\alpha=3/2$. Our treatment differs slightly, 
in terms of order $g^2$, from that of \cite{Boyanovsky:1995b},
where $\alpha=5/2$. 

In our numerical computation we just find the exponential 
damping (\ref{exp_damp}) with a value of $\Gamma$ which agrees with the 
theoretical expectation (\ref{gamma}).
The power behavior (\ref{pow_damp}) is apparently suppressed
due to a small coefficient.

\section{Some numerical results}
We have implemented numerically the formalism developed in the 
previous sections. We will discuss in this section some results of
our numerical simulations. We will pay special attention to
the phenomenon of Pauli blocking invoked 
in \cite{Boyanovsky:1995b}.
We have already seen that the occupation number cannot
exceed unity on account of the unitary evolution of the
mode functions $U_{\bfp,s}(t),V_{\bfp,s}(t)$. So an unlimited
parametric resonance cannot develop. A priori this should
not limit the production of particles as the available phase space 
is large. However, by a phenomenon similar to parametric resonance,
the production of particles turns out to be concentrated
within a very small band and it is only in this
resonant region where Pauli blocking 
can be effective. One should keep in mind, however, that even
for the bosonic case particle production shuts off in the
one-loop approximation \cite{deVega:1997}.

We expect Pauli blocking to be especially effective if the initial
amplitude of the inflaton field is large. A typical case is
displayed in Figs. 1-5. It corresponds to the parameters
$M=1,\lambda=1,g=2$, and $\phi(0)=0.6$. $\dot \phi (0)$ is taken
to be zero in all examples. We show the behavior of the inflaton
amplitude in Fig. 1, the conserved total energy and its classical and
fluctuation parts in Fig. 2 and the pressure in Fig. 3.
The time dependence of the particle number is displayed
in Fig. 4, using both definitions: $\caln_0(t)$, referring to quanta
of mass $m_0$ and $\caln_1(t)$ referring to quanta of mass
$m(t)$. The latter one is seen to behave more smoothly.
The momentum spectrum of the
occupation number $\caln^1_{p,s}(t)$ varies strongly with time. 
We display therefore, in Fig. 5, the envelope 
 obtained by selecting 
the maximal occcupation number reached at fixed momentum,
 as a function of $E_0$. The structure of 
this envelope shows  resonance-like enhancements
at threshold $E_0=1.2$, and at $E_0=1.9$. The maximal occupation number is
reached only at these two values. On the one hand there is no
Pauli blocking in the sense that all levels would be maximally 
occupied, on the other hand the unitary evolution does not 
allow for a parametric resonance with high occupation numbers.  
Clearly, for this parameter set the classical field is not able
to efficiently transfer energy into the fermion fluctuations 
and its amplitude stays essentially constant. These results are
similar to those obtained in \cite{Boyanovsky:1995b}.

The situation is simple in the case of very small excitations
and moderate couplings. In this case the scalar field can decay
into fermion-antifermion pairs. An exponential decrease is found
for the exact quantum evolution and for the linearized
equations of motion. This is displayed in Figs. 6-10.
Fig. 6 shows the exact evolution of $\phi(t)$ for the parameter set
$M=1, g=2, \lambda=1$, and $\phi(0)=0.01$. The amplitude is seen
to decrease exponentially, the decay rate being given approximately
by $\Gamma\simeq g^2/8\pi$. In the same Figure we also plot the
solution of the linearized equation of motion for which $\Gamma$ is
exactly equal to $g^2/8\pi$. The energy is transferred completely to the
quantum fluctuations. The pressure, plotted in Fig. 7, becomes 
asymptotically equal to one third of the total energy density
$\cale = 7.5 \cdot 10^{-5}$. The quantum ensemble created
is ultrarelativistic, as to be expected for massless quanta.
The momentum distribution of the occupation number is shown
in Fig. 8, it is characterized again by a resonance-like band.
The total particle number is plotted in Fig. 9.

So far the results correspond to the expectations. The situation
is however not as transparent. For intermediate initial
amplitudes $\phi(0)$ the relaxation can  shut off
even for case $2 m(0) < M$ where the scalar field can decay.
An example is given in Fig. 10, with the parameters
$M=1, g=1,\lambda=0$, and $\phi(0)=0.1$. The full evolution stagnates,
the linearized equations of motion show the expected
exponential decrease. 
On the other hand, even for large initial amplitudes the transfer
of energy can be as efficient as for scalar fields.
An example is displayed in Figs. 11-13. The amplitude $\phi(t)$,
plotted in Fig. 11, decreases by roughly a factor $3$. The
energy transfer is evident from Fig. 12. The spectrum of the
occupation number, presented in Fig. 13, shows that 
the resonance type band is occupied more strongly than in the first
example, which had a smaller initial amplitude, and otherwise the
same parameters. Obviously, the concept of Pauli-blocking is too
simple to describe the situation in an adequate way. If one
aims at a better  
understanding of the quantum evolution, analytical methods should 
be developed. The simplest approach could be an analysis
of the differential equation of the mode functions
for a given oscillating classical field, analogous to the
analysis of the Mathieu or Lam\'e equations.
In order to illustrate the behavior for a given oscillating
field we have solved numerically Eq. (\ref{fsec}) with
$m(t)=m_0 \cos t$ for various values of $m_0$ and
momentum $p$. We plot in Fig. 14
the envelope of the occupation number $\caln_{\rm env}$, i.e.,
the maximal occupation reached at fixed $p$, as a function of 
$E_0(p)$. The structure resembles the envelopes plotted in
Figs. 5 and 8. 

Finally, we shall present an example where we include
quantum fluctuations of both  the fermion field
and  the scalar field itself.
In Fig. 15 we display the behavior of the amplitude for 
this combined system, the parameters are
$M=1,g=2,\lambda=4$, and $\phi(0)=1$. The field is seen to relax
efficiently. On the contrary, if the scalar fluctuations are 
not included, the relaxation induced by the fermionic fluctuations
is small, as seen in Fig. 16. Fig. 17 shows the relaxation
for the case that only the scalar fluctuations are included, it is 
seen to start later and to be less efficient 
than for the combined system.
The growth of the fermionic and bosonic energy density is 
plotted in Fig. 18. The fermionic energy density
is smaller but rises earlier. Its asymptotic value is only
by roughly $20 \%$ higher than in the purely fermionic evolution.
 The fermion 
fluctuations seemingly act here as a kind of catalysator supporting
the development of the  bosonic quantum fluctuations.
 
\section{Conclusion}
We have developed the quantum field theory
of the out-of-equilibrium evolution of
fermionic quantum fluctuations driven by a scalar field. 
The quantum back-reaction has been taken into account in 
one-loop approximation. We have formulated the renormalization
of the equations of motion and of the energy-momentum tensor
in a covariant form and independent of the initial conditions.
A restriction of suitable initial conditions for the
fermionic quantum system, as required by 
the removal of initial singularities, has been 
obtained by selecting a Fock space built on a Bogoliubov-transformed
vacuum state. Furthermore, we have formulated the renormalized equations
for the case of a spatially flat FRW metric.

We have numerically implemented the evolution equations
and we have presented some examples for the evolution of the
quantum system. If the initial amplitude of the scalar field
is very small the system evolves as predicted for the
linearized equations of motion, formulated in section 7.
It can be described as a decay of the scalar field into
fermion-antifermion pairs. If the initial amplitude 
is larger, the evolution depends on the way in which
a kind of resonance band at low momenta is situated kinematically
and how it is filled. In some cases the fermions are indeed 
ineffective in damping the oscillation of the 
classical field, in others the relaxation develops as
in the bosonic case and a considerable part of
the energy is transferred to the quantum fluctuations.
Analytical studies should help to clarify the features 
observed for large-amplitude oscillations.  
An example where bosonic as well as fermionic fluctuations are
included shows an interesting interplay where the fermions
catalyze  the development of bosonic fluctuations.

We think that these results show that nonequilibrium
systems with fermionic fluctuations
show more interesting features and may play a more interesting 
r\^ole in cosmology than previously assumed.

\newpage 
\setcounter{equation}{0}

\begin{appendix}
\section{Perturbative expansion of the mode functions}
\setcounter{equation}{0}
We have introduced in sections 2 and 3 the mode functions
$f_p(t)$ and $h_p(t)$ which are related via Eq. (\ref{h_def}).
In this Appendix we will analyse the perturbation expansion
and ultraviolet behavior of the functions $h_p(t)$.
These mode functions satisfy the differential equation
\be\label{fdiffeq}
\ddot{h}_p-2iE_0\dot{h}_p=-V(t)\left[1+h_p\right]\kma
\ee
with the initial conditions $h_p(0)=\dot{h}_p(0)=0$.
We expand  $h_p$ with respect to orders in $V(t)$
by writing
\be
\label{entwicklung}
h_p= h_p^{(1)}+h_p^{(2)}+h_p^{(3)}+ .... \kma
\ee
where $h_p^{(n)}(t)$ is of n'th order in $V(t)$.
$h_p^{\overline{(n)}}$
denotes the sum over all orders beginning with the n'th one. 
\be
h_p^{\overline{(n)}}=\sum_{l=n}^\infty h_p^{(n)}
\kma\ee
so that
\be
h_p\equiv h_p^{\overline{(1)}}
=h_p^{(1)}+h_p^{\overline{(2)}}\pkt
\ee
The integral equation for the function $h_p$ can be derived
in a straightforward way from the 
differential equation satisfied by the functions $f_p$;
it reads
\be
h_p=\frac{i}{2E_0}\int_0^t dt' \left(e^{2iE_0(t-t') }-1\right)
V(t')\left[1+h_p(t')\right]\pkt
\ee
Using this integral equation we can obtain the functions
$h^{(n)}_p(t)$ by iteration \cite{Baacke:1997a}. $h_p^{(1)}$ is
given by
\be
h_p^{(1)}=\frac{i}{2E_0}\int_0^t dt' 
\left(e^{2iE_0(t-t') }-1\right)V(t')\pkt
\ee
Using integrations by parts 
this function can be analyzed with respect to orders in
$E_0$ via
\bea \nonumber
h_p^{(1)}&=&\frac{-i}{2E_0}\int_0^t dt' V(t')
+\sum_{l=0}^{n-1}\left(\frac{-i}{2E_0}\right)^{l+2}
\left[V^{(l)}(t)-e^{2iE_0 t} V^{(l)}(0)\right]\\
&&+ \left(h_p^{(1)}\right)_{\overline{n}}
\eea
with
\be
\left(h_p^{(1)}\right)_{\overline{n}}
= -\left(\frac{-i}{2E_0}\right)^{n+1}
\int_0^t dt'e^{2iE_0(t-t')}
V^{(n)}(t') \kma \label{hexp} 
\ee
Here $V^{(l)}(t)$ denotes the $l$th derivative of $V(t)$; the
subscript $\overline{n}$ indicates that 
the expression in parentheses has been reduced to
negative powers of $E_0$ equal or higher than $n$.
For energy density and pressure we need the expansion of
$\dot h_p^{(1)}(t)$ as well. From Eq. (\ref{hexp}) and the relation
\be
\dot h_p^{(1)}=2 i E_0 h_p^{(1)}- \int _0^tdt'V(t')
\ee
 we find
\bea \nonumber
\dot h_p^{(1)}&=&\sum_{l=0}^n \left(\frac{-i}{2E_0}\right)^{l+1}
\left[V^{(l)}(t)-e^{2iE_0 t} V^{(l)}(0)\right]\\
&&-\left(\frac{-i}{2E_0}\right)^{n+1}
\int_0^t dt'e^{2iE_0(t-t')}
V^{(n+1)}(t') \pkt \label{hdotexp} 
\eea
Similar expressions hold for the higher $h_p^{(n)}$
and $\dot h_p^{(n)}$.

In the numerical implementation the functions 
$\left(h_p^{(1)}\right)_{\overline{n}}$ can be obtained 
as the Fourier transform of the $n$'th derivative of $V(t)$. 
Its computation need just one update per time step. 
Alternatively one may construct
differential equations satisfied by these functions, e.g.,
\be
\left(\ddot h^{(1)}_p\right)_{\overline{ 2}}
- 2 i E_0\left(\dot h^{(1)}_p\right)_{\overline{ 2}}=
 \frac{i}{2E_0}\dot V \pkt
\ee

In order to isolate the divergent terms in the fluctuation integrals
of the equation of motion and of the energy momentum tensor we
need an expansion of the mode function up to order $O\left[(E_0)^{-3}\right]$
and $O\left[(E_0)^{-4}\right]$, respectively.
For this reason we will give relevant expansions of the $h_p$ in the following.

The expansion of $h_p^{(1)}$ up to $O\left[(E_0)^{-3}\right]$ gives 
\bea
h_p^{(1)}&=&-\frac{i}{2E_0}\il{}{'}V(t')-\frac{V(t)}{4(E_0)^2}+
\frac{i\dot{V}(t)}{8(E_0)^3}
-\frac{i\dot{V}(0)}{8(E_0)^3}e^{2iE_0t}
\nonumber\\&&
-\frac{i}{8(E_0)^3}\il{}{'}e^{2iE_0(t-t')}\ddot{V}(t)
\eea
For $h_p^{(2)}$ we obtain
\bea
h_p^{(2)}&=&-\frac{1}{8(E_0)^2}\left[\il{}{'}V(t')\right]^2+
\frac{i}{8(E_0)^3}\il{}{'}V^2(t')
\nonumber\\&&+\frac{i}{8(E_0)^3}V(t)\il{}{'}V(t')+ 
\left(h^{(2)}_p\right)_{\overline{4}}
\eea
where $\left(h^{(2)}_p\right)_{\overline{4}}$ includes 
all terms of $h_p^{(2)}$ that have
at least four negative powers of $E_0$.
It satisfies the differential equation
\bea
\left(\ddot{h}^{(2)}_p\right)_{\overline{4}}- 
2iE_0\left(\dot{h}^{(2)}_p\right)_{\overline{4}}&=&
-V(t)\left(h^{(1)}_p\right)_{\overline{3}}-\frac{5i}{8(E_0)^3}\dot{V}(t)V(t)
\nonumber \\&&-\frac{i}{8(E_0)^3}\ddot{V}(t)\il{}{'}V(t')\pkt
\eea
Finally, via integration by parts, $h_p^{(3)}$ takes the form
\be
h_p^{(3)}=\frac{i}{48(E_0)^3}\left[
\il{}{'}V(t')
\right]^3+ \left(h^{(3)}_p\right)_{\overline{4}}
\kma \ee
where the last term satifies the differential equation
\bea
\left(\ddot{h}^{(3)}_p\right)_{\overline{4}}- 
2iE_0\left(\dot{h}^{(3)}_p\right)_{\overline{4}}&=&
-V(t)\left(h^{(2)}_p\right)_{\overline{3}}-
\frac{i\dot{V}(t)}{16(E_0)^3}\il{}{'}V(t')
\nonumber \\ &&-\frac{iV^2(t)}{8(E_0)^3}\il{}{'}V(t')
\pkt \eea
For the integrands of the fluctuation integral we need
repeatedly the two expressions:
\be
\cali_1(p,t)=2 \re h_p^{\overline{(1)}}+|h_p^{\overline{(1)}}|^2
\ee
and
\be
\cali_2(p,t)=2\im\dot{h}_p^{\overline{({1})}}
+2\im h_p^{\overline{(1)}*}\dot{h}_p^{\overline{({1})}}
\pkt\ee
Using the expansion up to $O\left[(E_0)^{-3}\right]$ we obtain
\bea
\cali_1(p,t)&=&\frac{1}{E_0}\il{}{'}\im V(t')
-\frac{\re V(t)}{2(E_0)^2}+
\frac{1}{2(E_0)^2}\left[\il{}{'}\im V(t')\right]^2\nonumber\\
&&-\frac{\im \dot{V}(t)}{4(E_0)^3}+
\frac{\im \dot{V}(0)}{4(E_0)^3}\cos\left(2E_0t\right)
+\frac{\re \dot{V}(0)}{4(E_0)^3}\sin\left(2E_0t\right)\nonumber\\
&&-\frac{1}{4(E_0)^3}\il{}{'}\im V^2(t')+
\frac{1}{6(E_0)^3}\left[\il{}{'}\im V(t')\right]^3
\\\nonumber
&&-\frac{1}{2(E_0)^3}\re V(t)\il{}{'}\im V(t')
+ 2\re \left(h^{\overline{(1)}}_p\right)_{\overline{4}}
+\left(|h^{\overline{(1)}}_p|^2
\right)_{\overline{4}}\kma
\eea
where we have introduced the terms of $O\left[(E_0)^{-4}\right]$ as
\be 2\re \left(h^{\overline{(1)}}_p\right)_{\overline{4}}
=2\re \left(h^{(1)}_p+h^{(2)}_p+h^{(3)}_p\right)
_{\overline{4}}+ 2\re h^{\overline{(4)}}_p
\ee
and
\be
\left(|h^{\overline{(1)}}_p|^2\right)_{\overline{4}}
=\left(|h^{{(1)}}_p|^2\right)_{\overline{4}}
+2\re\left(h^{(1)}_p h^{(2)*}_p\right)_{\overline{4}}+|h^{\overline{(2)}}_p|^2
+2\re\left(h^{(1)}_p h^{\overline{(3)}*}_p\right)
\pkt
\ee
Inserting the real and imaginary parts of the 
potential $V(t)$ we finally have the result
\bea
\cali_1(p,t)&=&-\frac{1}{E_0}\left[m(t)-m(0)\right]
-\frac{1}{(E_0)^2}\left[m(t)m(0)-m^2(0)\right]\nonumber\\
&&
+\frac{1}{(E_0)^3}\left[
\frac 1 2 m^3(t)+m^3(0)-\frac 3 2 m(t)m^2(0)\right.
\nonumber\\
&&\hspace{1,5cm}\left.
-\frac{1}{4}\ddot{m}(0)\cos \left(2E_0t\right)
+\frac{1}{2}\dot{m}(0)m_0\sin \left(2E_0t \right)
\right]\nonumber\\
&&+ 2\re \left(h^{\overline{(1)}}_p\right)_{\overline{4}}
+\left(|h^{\overline{(1)}}_p|^2
\right)_{\overline{4}}
\kma \\
\cali_2(p,t)&=&
-\frac{\re V(t)}{E_0}-\frac{\im \dot{V}(t)}{2(E_0)^2}
+\frac{\im \dot{V}(0)}{(2E_0)^2}\cos(2E_0t)
\nonumber\\
&&+\frac{\re \dot{V}(t)}{2(E_0)^2}\sin(2E_0t)
+\frac{\re \ddot{V}(t)}{4(E_0)^3}
-\frac{\re \ddot{V}(0)}{4(E_0)^3}\cos(2E_0t)\nonumber\\
&&+\frac{\im \ddot{V}(0)}{4(E_0)^3}\sin (2E_0t)
+\frac{1}{(E_0)^2}\re V(t)\il{}{'}\im V(t')
\nonumber
\\
&&
+\frac{3\left[\re V(t)\right]^2}{4(E_0)^3}
-\frac{\left[\im V(t)\right]^2}{4(E_0)^3}
-\frac{\im \dot{V}(t)}{2(E_0)^3}\il{}{'}\im V(t')\nonumber\\
&&
-\frac{\re V(t)}{2(E_0)^3}\left[\il{}{'}\im V(t')\right]^2+
\frac{\re \dot{V}(0)}{2(E_0)^3}\cos(2E_0t)\il{}{'}\re V(t')\nonumber\\
&&-\frac{\im \dot{V}(0)}{2(E_0)^3}\sin(2E_0t)\il{}{'}\re V(t')\nonumber\\
&&+2\im\left( \dot{h}^{\overline{(1)}}_p\right)_{\overline{4}}
+2\im\left( h_p^{*\overline{(1)}}
\dot{h}^{\overline{(1)}}_p\right)_{\overline{4}}
\pkt \eea
For completeness we also give the expansion up to ${O}\left[(E_0)^{-4}\right]$:
\bea
\cali_1(p,t)
&=&\frac{1}{E_0}\il{}{'}\im V(t')
-\frac{\re V(t)}{2(E_0)^2}+\frac{1}{2(E_0)^2}
\left[\il{}{'}\im V(t')\right]^2\nonumber\\
&&-\frac{\im \dot{V}(t)}{4(E_0)^3}+
\frac{\im \dot{V}(0)}{4(E_0)^3}\cos\left(2E_0t\right)
+\frac{\re \dot{V}(0)}{4(E_0)^3}\sin\left(2E_0t\right)\nonumber\\
&&-\frac{1}{4(E_0)^3}\il{}{'}\im V^2(t')+
\frac{1}{6(E_0)^3}\left[\il{}{'}\im V(t')\right]^3
\nonumber\\
&&-\frac{1}{2(E_0)^3}\re V(t)\il{}{'}\im V(t')
+\frac{3\left[\re V(t)\right]^2}{8(E_0)^4}-
\frac{\left[\im V(t)\right]^2}{4(E_0)^4}\nonumber
\\
&&
-\frac{\im \dot{V}(t')}{4(E_0)^4}\il{}{'}\im V(t')-
\frac{1}{4(E_0)^4}\il{}{'}\im V(t')\il{}{'}
\im V^2(t')\nonumber\\
&&-\frac{1}{4(E_0)^4}\re V(t)\left[\il{}{'}\im V(t')\right]^2
+\frac{1}{24(E_0)^4}\left[\il{}{'}\im V(t')\right]^4\nonumber\\&&
+\frac{\re \dot{V}(0)}{4(E_0)^4}\cos(2E_0t)\il{}{'}
\re V(t')+\frac{\re \ddot{V}(t)}{8(E_0)^4}\nonumber\\&&
-\frac{1}{4(E_0)^4}\im \dot{V}(0)\sin(2E_0t)\il{}{}\re V(t')
-\frac{\re \ddot{V}(0)}{8(E_0)^4}\cos(2E_0t)\nonumber\\
&&+\frac{\im \ddot{V}(0)}{8(E_0)^4}\sin(2E_0t)
+ 2\re \left(h^{\overline{(1)}}_p\right)_{\overline{5}}+
\left(|h^{\overline{(1)}}_p|^2
\right)_{\overline{5}}\kma
\eea
where
\be 2\re \left(h^{\overline{(1)}}_p\right)_{\overline{5}}=
2\re \left(h^{(1)}_p+h^{(2)}_p+h^{(3)}_p
+h^{(4)}_p
\right)
_{\overline{5}}+ 2\re h^{\overline{(5)}}_p
\pkt \ee
In terms of the notation introduced above the 
alternative representation for the functions
$K_{\rm F}, K_{\rm E}$, and $K_{\rm P}$ become now
\bea
K_{\rm F}(p,t)&=&-\left[\cali_1(p,t)\right]_{\overline{4}}
+\frac{m_0}{E_0}
\left[\cali_1(p,t)\right]_{\overline{3}}
\ko \\
K_{\rm E}(p,t)&=&-E_0
\left[\cali_1(p,t)\right]_{\overline{5}}
+\left[\cali_2(p,t)\right]_{\overline{4}}
+m_0\left[\cali_1(p,t)\right]_{\overline {4}}
\\ \nonumber&&-\frac{m_0}{E_0}\left[\cali_2(p,t)\right]_{\overline{3}}
\kma \\
K_{\rm P}(p,t)&=&-\frac 4 3 E_0
\left[\cali_1(p,t)\right]_{\overline{5}}
+\left(\frac 4 3 m_0-\frac 1 3 m(t)\right)
\left[\cali_1(p,t)\right]_{\overline{4}}
\\ \nonumber&&+\frac{m(t)m_0}{3E_0}\left[\cali_1(p,t)\right]_{\overline{3}}
+\frac 4 3 \left[\cali_2(p,t)\right]_{\overline{4}}
-\frac{4m_0}{3E_0}\left[\cali_2(p,t)\right]_{\overline{3}}
\pkt
\eea
Unlike the case of scalar fluctuations \cite{Baacke:1997a}, 
the evaluation of these expressions has required extensive
algebra, and the numerical implementation requires further 
efforts. This reflects the fact, that fermion loops are 
divergent up graphs with four external lines.
The definition of $K_{\rm F},K_{\rm E}$, and
$K_{\rm P}$ given in section 4 is much easier to handle,
it involves, however, potentially dangerous
differences between expressions which are computed numerically
and leading perturbative terms. We found in our numerical
computations that the simple subtraction was tolerable; this is due,
here, to the fact that most integrals are dominated by the low
momentum region. The analysis presented in this Appendix was
necessary in any case, however, in order to find the divergent contributions.


\section{Bogoliubov transformation}
\setcounter{equation}{0}
Here we recall the basic formulae for the Bogoliubov transformation
of spin $1/2$ fields (see e.g. \cite{Bogoliubov:1983}).
A general canonical transformation compatible with 
the anticommutation relations is given, up to a further trivial
phase transformation, by
\bea
b_{\bfp,s}&=&\cos (\beta_{\bfp,s}) \tilde b_{\bfp,s}+
\sin(\beta_{\bfp,s})\exp^{i\delta_{\bfp,s}}\tilde d^\dagger_{-\bfp,s}
\\
d^\dagger_{-\bfp,s}&=&
-\sin(\beta_{\bfp,s})e^{-i\delta_{\bfp,s}}\tilde b_{\bfp,s}
+\cos (\beta_{\bfp,s}) \tilde d^\dagger_{-\bfp,s}
\pkt\eea
The state annihilated by the new operators $\tilde b_{\bfp,s}$
and $\tilde d_{\bfp,s}$ is given in discrete notation by
\be
|\tilde 0 \rangle = \prod_{\bfp,s}
\{ \cos (\beta_{\bfp,s}) +\frac{1}{2 E_0(p) V}
\sin(\beta_{\bfp,s})e^{-i\delta_{\bfp,s}}
d^\dagger_{-\bfp,s}b^\dagger_{\bfp,s}\} |0\rangle
\pkt\ee
For the expectation values of the various bilinear products
of interest here one finds, in continuum notation,
\bea
\langle \tilde 0|b^\dagger_{\bfp,s}b_{\bfp',s'}|\tilde 0\rangle
&=&(2\pi)^3 2 E_0\delta_{ss'}\delta^3(\bfp-\bfp')\sin^2\beta_{\bfp,s}
\kma \\
\langle \tilde 0|d_{-\bfp,s}d^\dagger_{-\bfp',s'}|\tilde 0\rangle
&=&(2\pi)^3 2 E_0\delta_{ss'}\delta^3(\bfp-\bfp')\cos^2\beta_{\bfp,s}
\kma \\
\langle \tilde 0|d_{-\bfp,s}b_{\bfp',s'}|\tilde 0\rangle
&=&(2\pi)^3 2 E_0\delta_{ss'}\delta^3(\bfp-\bfp')
e^{i\delta_{\bfp,s}}\cos\beta_{\bfp,s}\sin\beta_{\bfp,s}
\kma \\
\langle \tilde 0|b^\dagger_{\bfp,s}d^\dagger_{-\bfp',s'}|\tilde 0\rangle
&=&(2\pi)^3 2 E_0\delta_{ss'}\delta^3(\bfp-\bfp')
e^{-i\delta_{\bfp,s}}\cos\beta_{\bfp,s}\sin\beta_{\bfp,s}
\kma \eea
so that the fluctuation  integral becomes
\bea
\tilde \calf(t)&=&\sum_s\inte\left\{\bar U_{\bfp, s} 
U_{\bfp,s}\sin^2\beta_{\bfp,s}
+\bar V_{-\bfp,s}V_{\-\bfp,s}\cos^2\beta_{\bfp,s}
\right.\nonumber\\
&&
\hspace{10mm}+\bar V_{-\bfp,s}U_{\bfp,s}e^{i\delta_{\bfp,s}}
\cos\beta_{\bfp,s}\sin\beta_{\bfp,s}\\
\nonumber 
&&
\hspace{10mm}\left.+\bar U_{\bfp,s} 
V_{-\bfp,s}e^{-i\delta_{\bfp,s}}\cos \beta_{\bfp,s}
\sin \beta_{\bfp,s}\right\}
\pkt\eea
This corresponds to replacing
\be
V_{-\bfp,s}(t) \Rightarrow \cos(\beta_{\bfp,s})V_{-\bfp,s}(t)+
e^{i\delta_{\bfp,s}}\sin(\beta_{\bfp,s})U_{\bfp,s}(t) 
\ee
in the original expression (\ref{f_int}). With the same
substitution one obtains the expectation values for the
energy and pressure. For the particle number the substitution is made
in the Bogoliubov coefficients, Eqs. (\ref{D_0}) 
and (\ref{D_1}).
As in the
case of scalar fluctuations \cite{Baacke:1998a},
we do not need the phase $\delta_{\bfp,s}$
as long as we restrict the initial condition for the
classical field to $\dot\phi(0)=0$.
\end{appendix}

\newpage

\section*{Figure Captions}

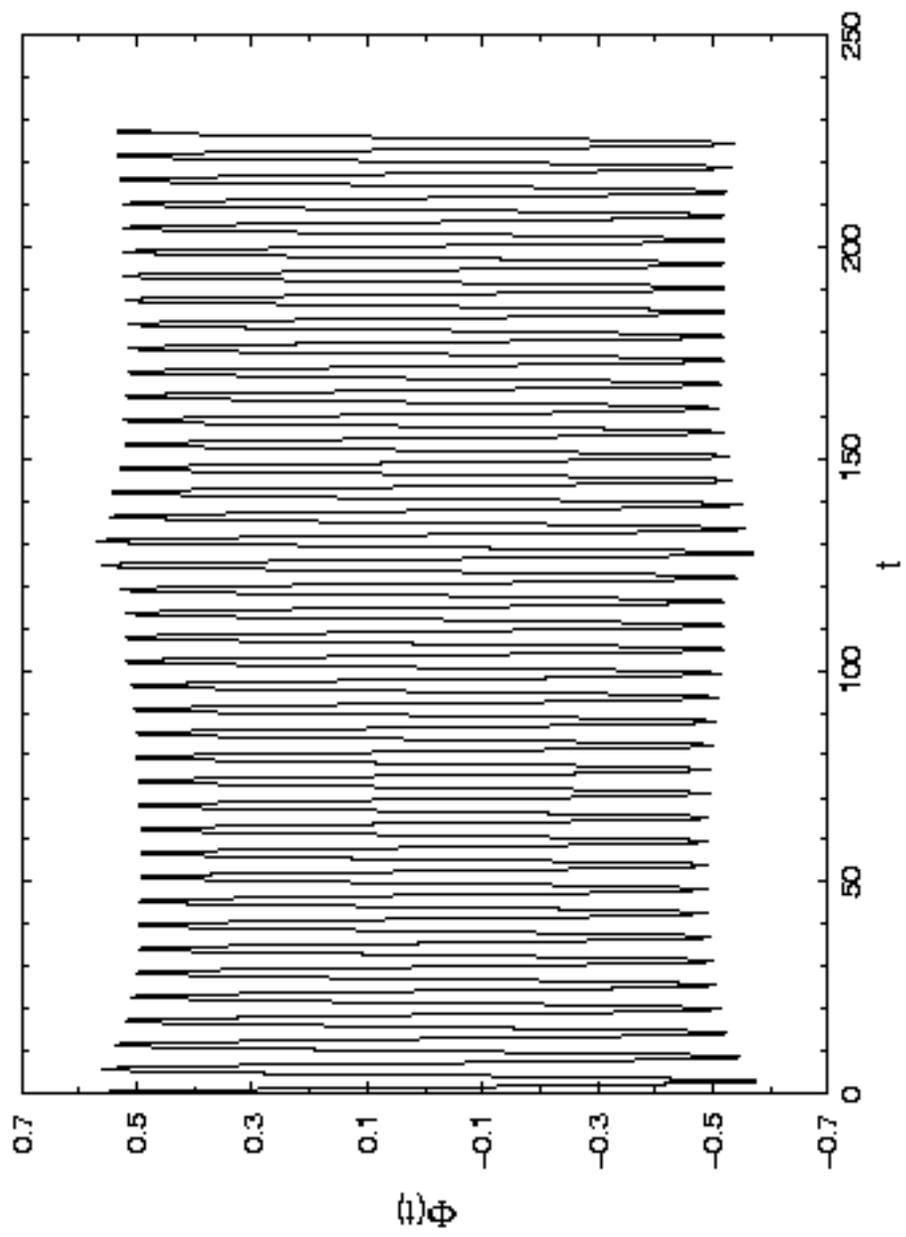
{\bf Fig. 1}: $\phi(t)$ for $M=1,g=2,\lambda=1$, 
and $\phi(0)=0.6,\dot\phi(0)=0$.
\\ \\
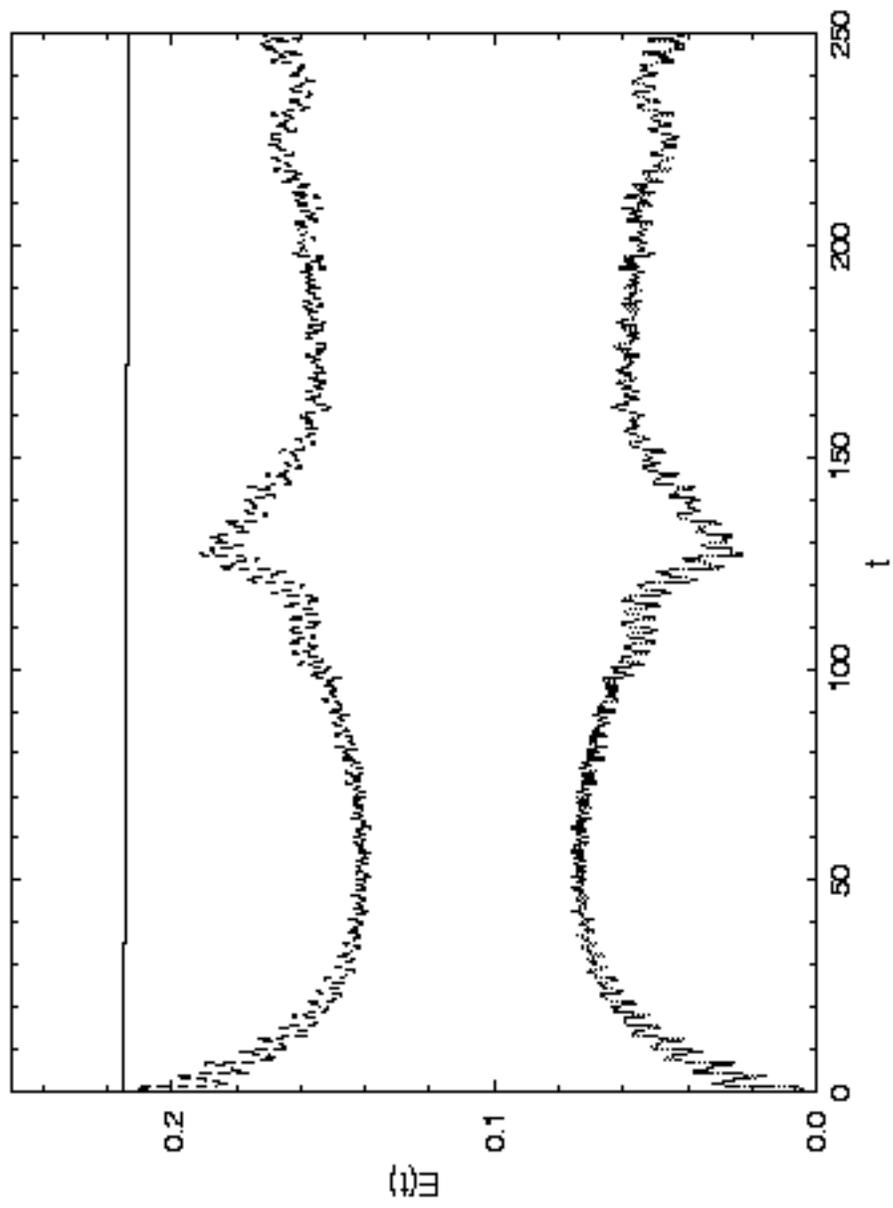
{\bf Fig. 2}: Classical energy (dashed line), 
fluctuation energy (dotted line), 
and total energy (solid line) for the same parameters as in Fig. 1.
\\ \\
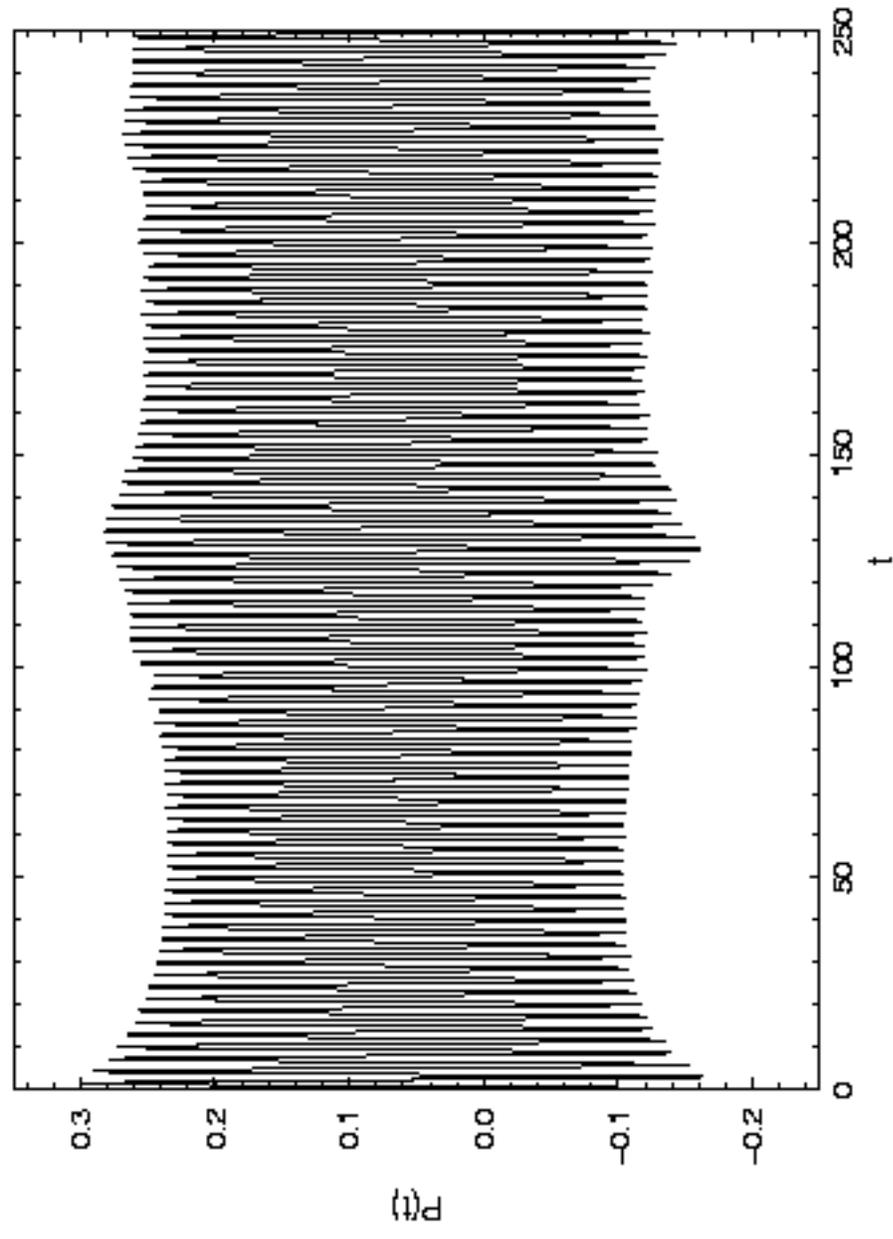
{\bf Fig. 3}: Total pressure for the same parameters as in Fig. 1.
\\ \\
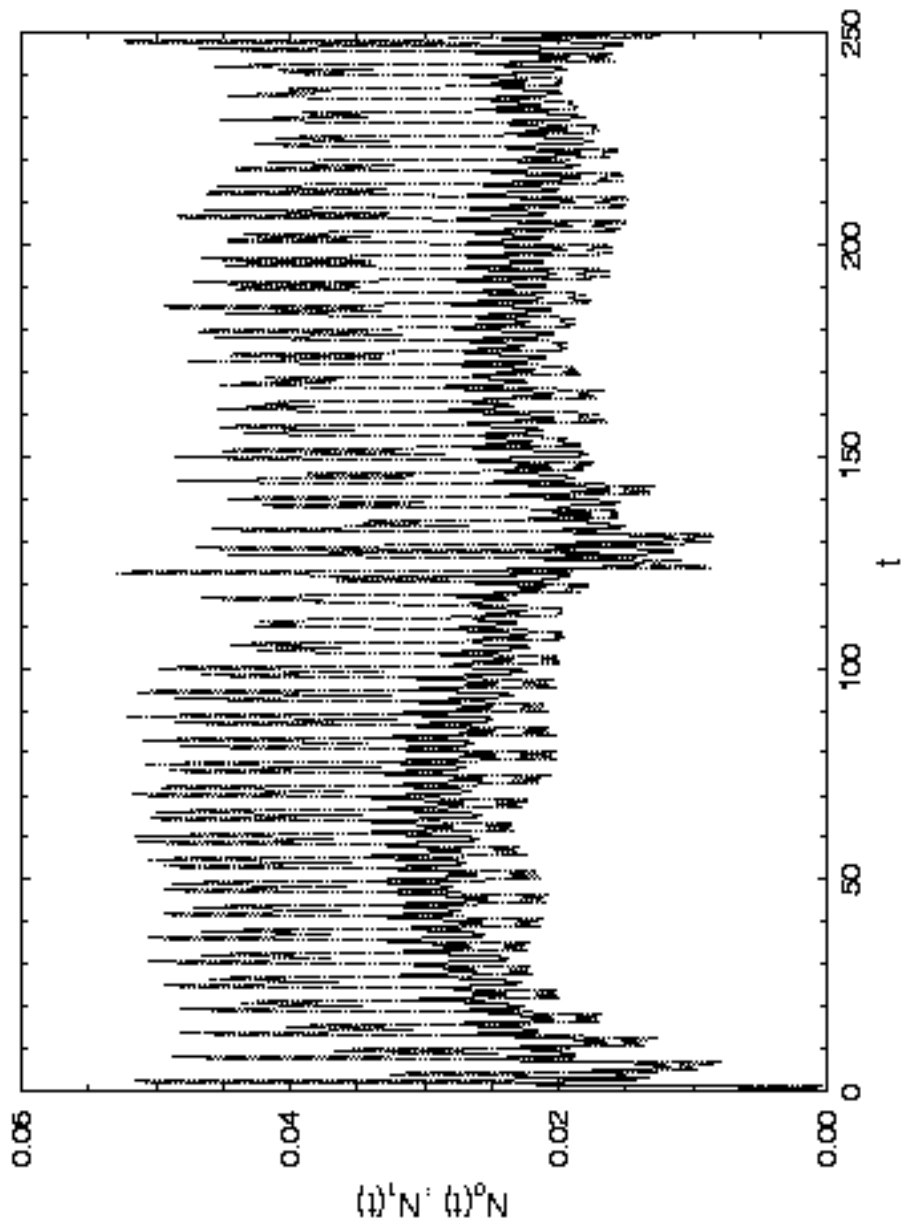
{\bf Fig. 4}: Total particle number $\caln_1(t)$ (solid line) 
and  $\caln_0(t)$ (dotted line) 
for the same parameters as in Fig. 1.
\\ \\
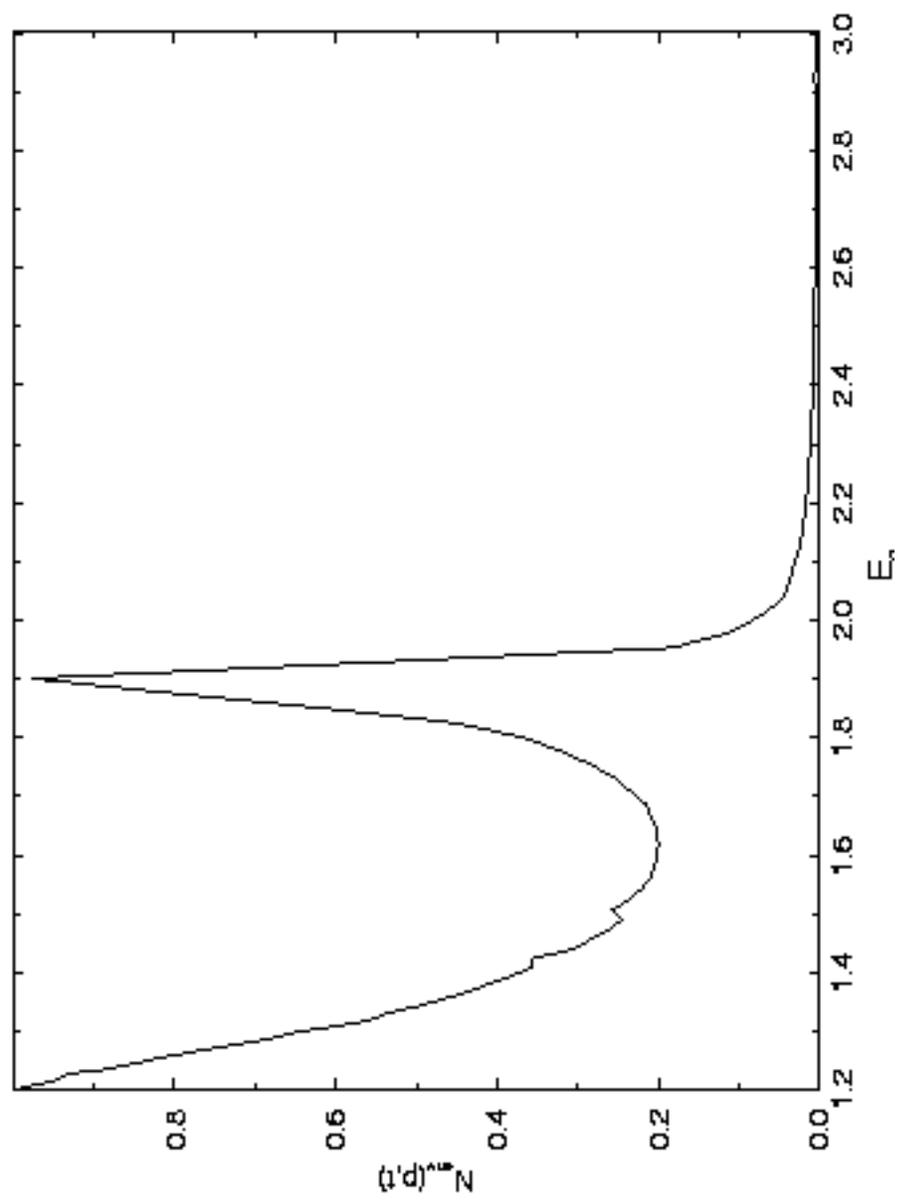
{\bf Fig. 5}: Maximal occupation number $\caln_{\rm env}$
 as a function of $E_0$ 
for the same parameters as in Fig. 1.
\\ \\
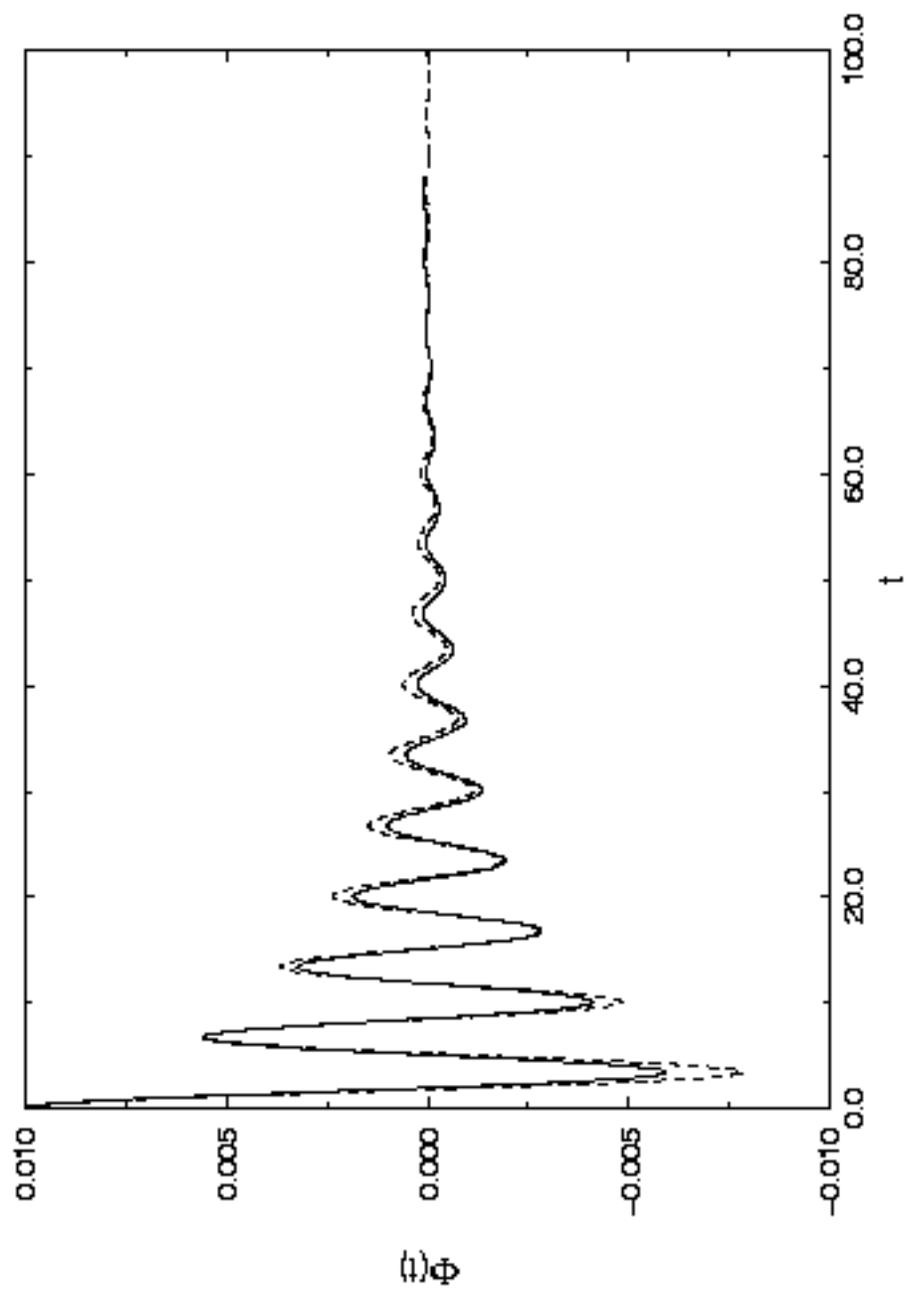
{\bf Fig. 6}: Exact quantum evolution (dashed line) and linearized evolution
(solid line) of $\phi(t)$ for $M=1,g=2,\lambda=1$, 
and $\phi(0)=0.01,\dot\phi(0)=0$.
\\ \\
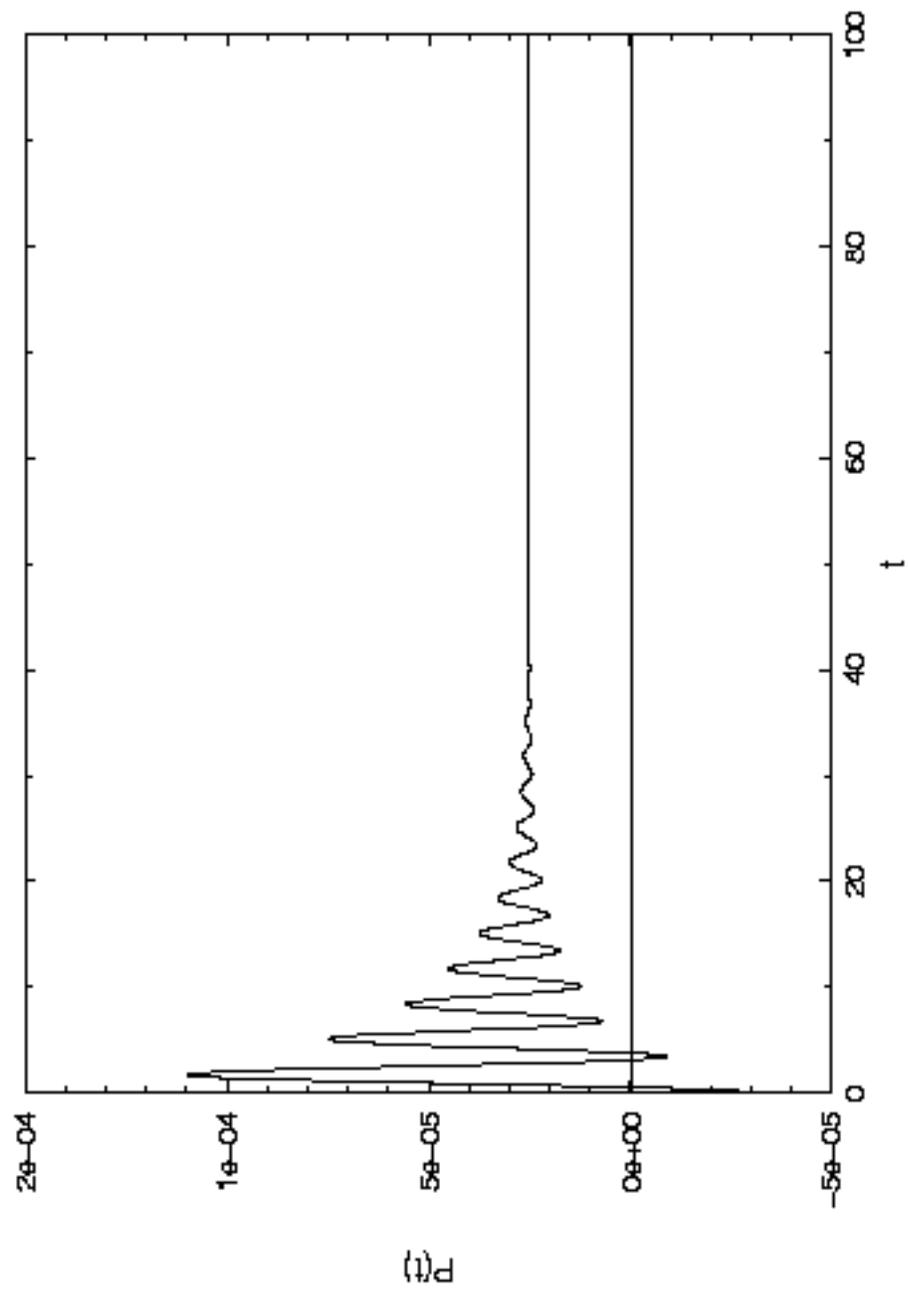
{\bf Fig. 7}: Total pressure for the exact evolution for the same 
parameters as in Fig. 6.
\\ \\
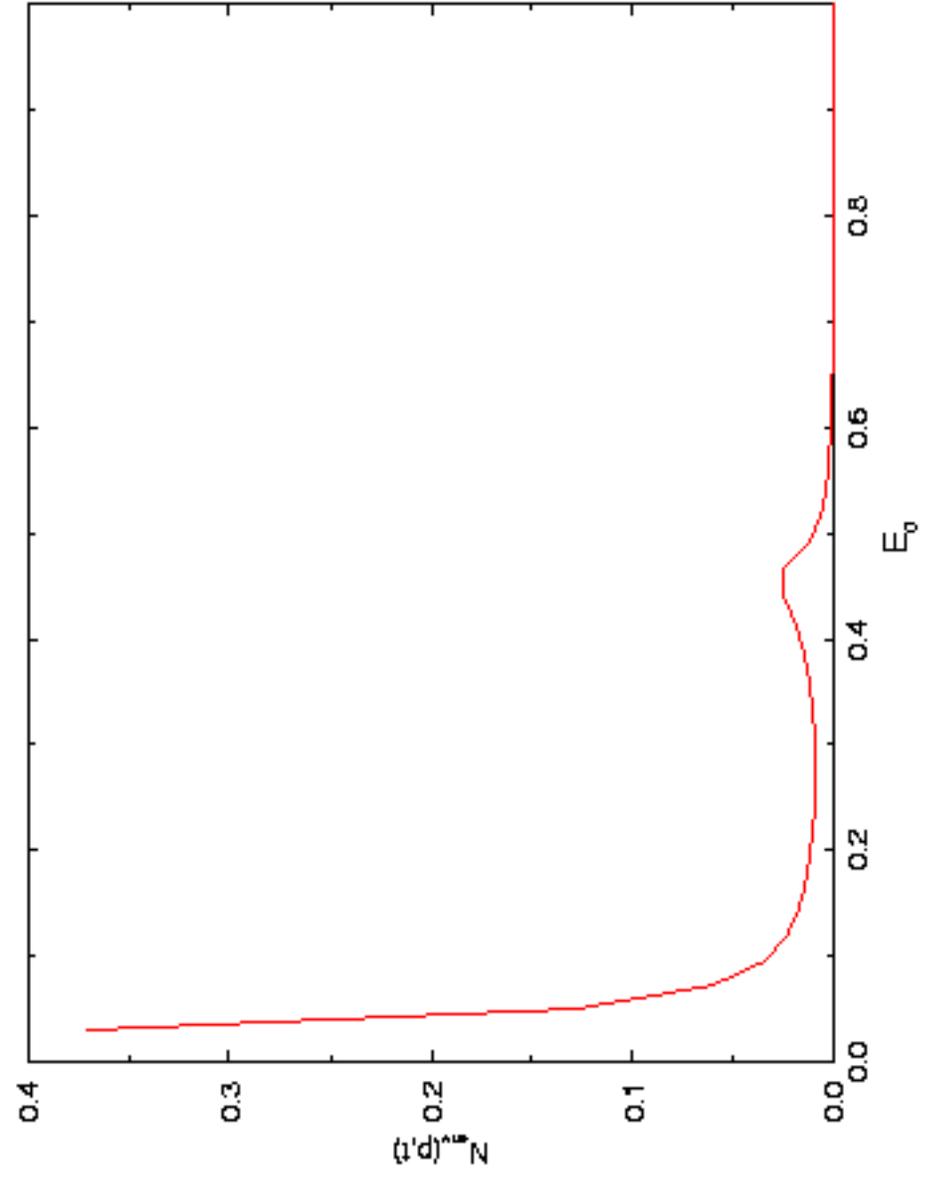
{\bf Fig. 8}: Maximal occupation number $\caln_{\rm env}$
 as a function of $E_0$ for the 
same parameters as in Fig. 6.
\\ \\
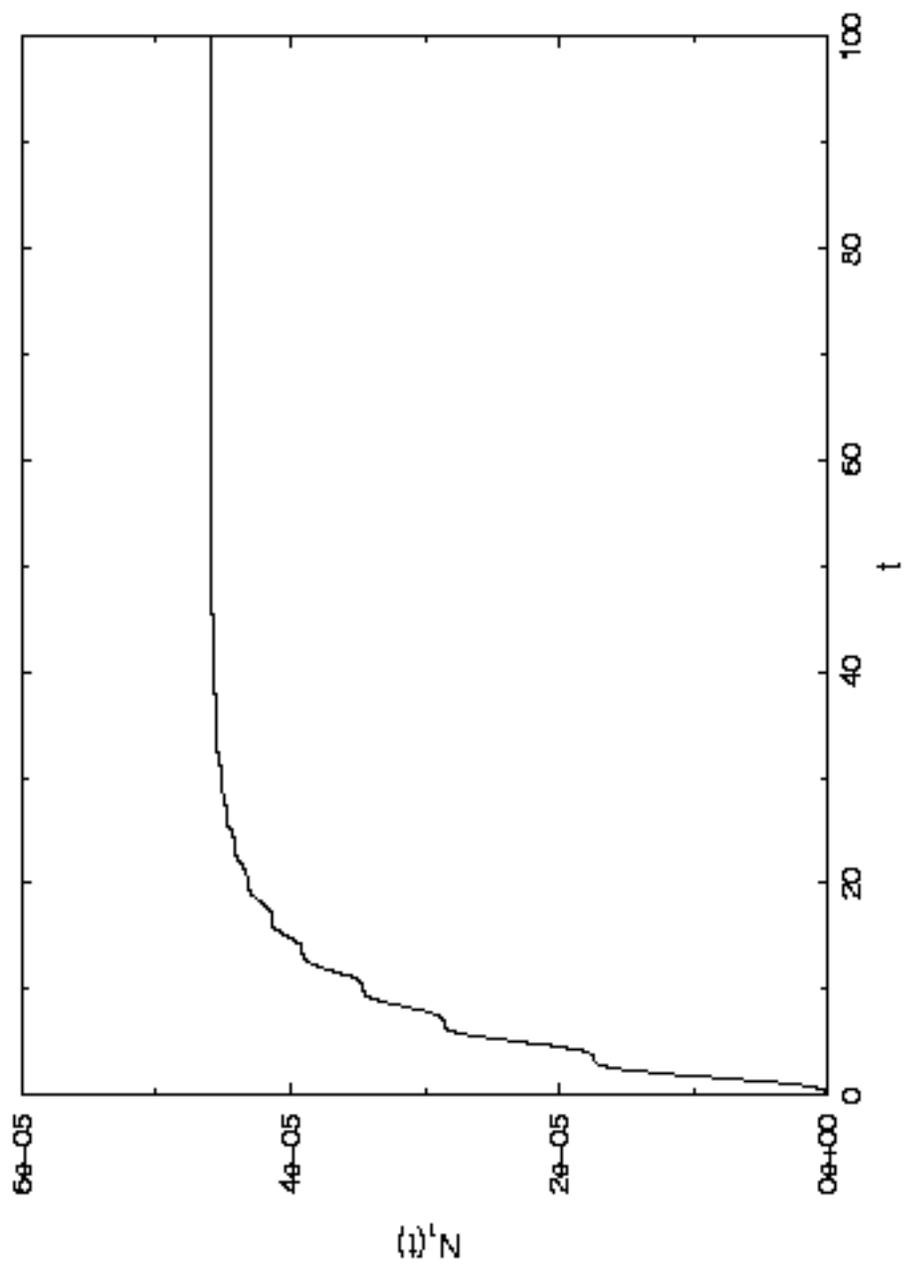
{\bf Fig. 9}: Total particle number $\caln_1(t)$ for the same case as in Fig. 6.
\\ \\
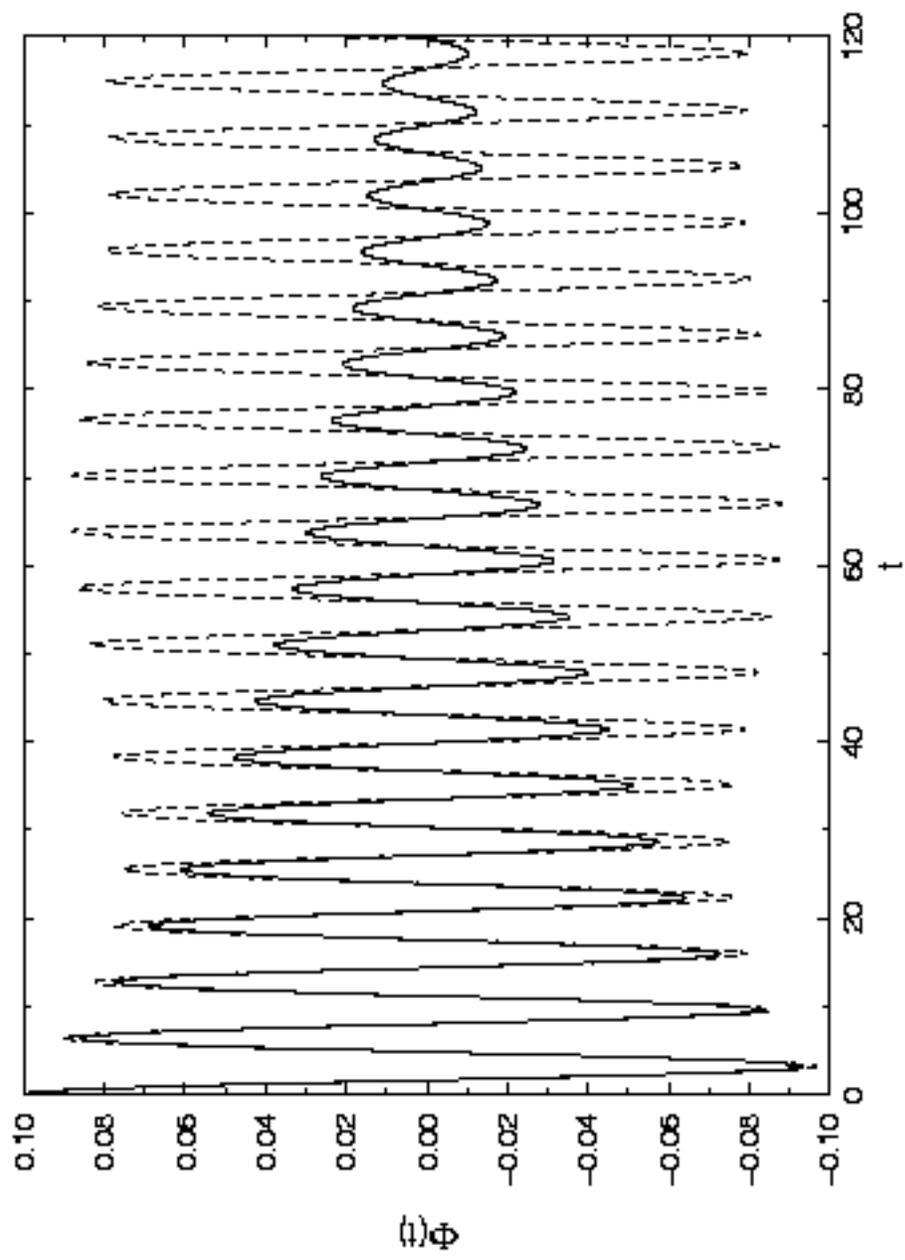
{\bf Fig. 10}: Exact quantum evolution (dashed line) and 
linearized evolution
(solid line) of $\phi(t)$ for $M=1,g=1,\lambda=0$ 
and $\phi(0)=0.1,\dot\phi(0)=0$.
\\ \\
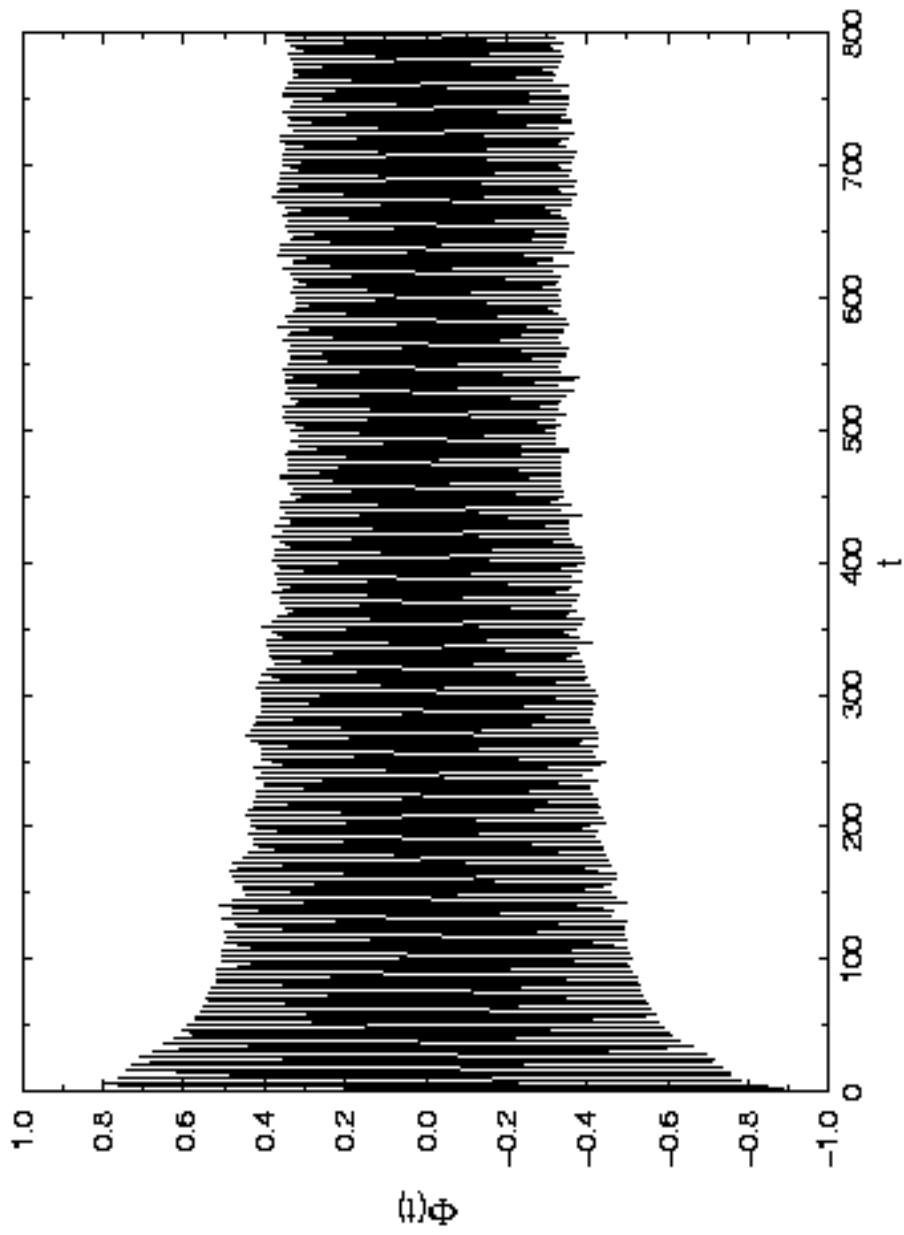
{\bf Fig. 11}: $\phi(t)$ for $M=1,g=2,\lambda=1$ and $\phi(0)=2,\dot\phi(0)=0$
\\ \\
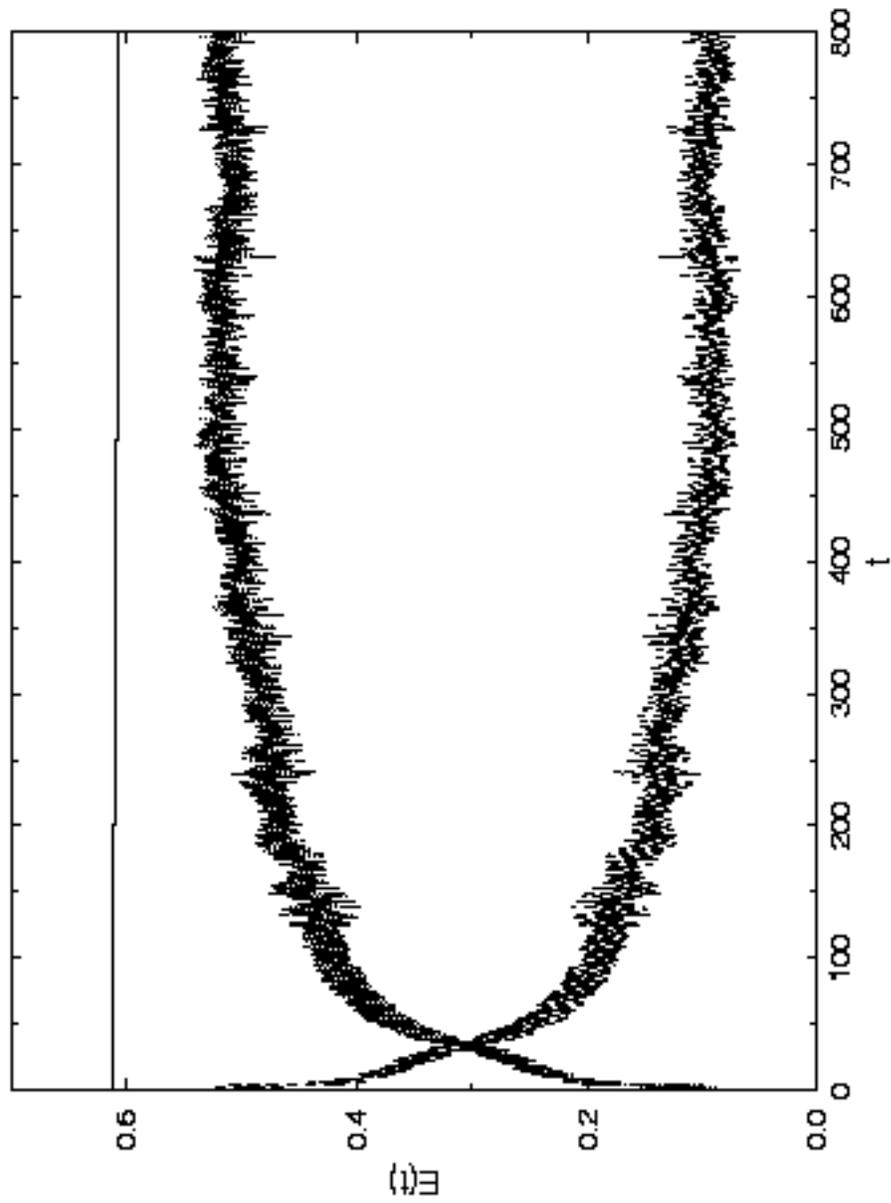
{\bf Fig. 12}: Classical energy (dashed line), 
fluctuation energy (dotted line), and 
total energy (solid line) for the same parameters as in Fig. 11.
\\ \\
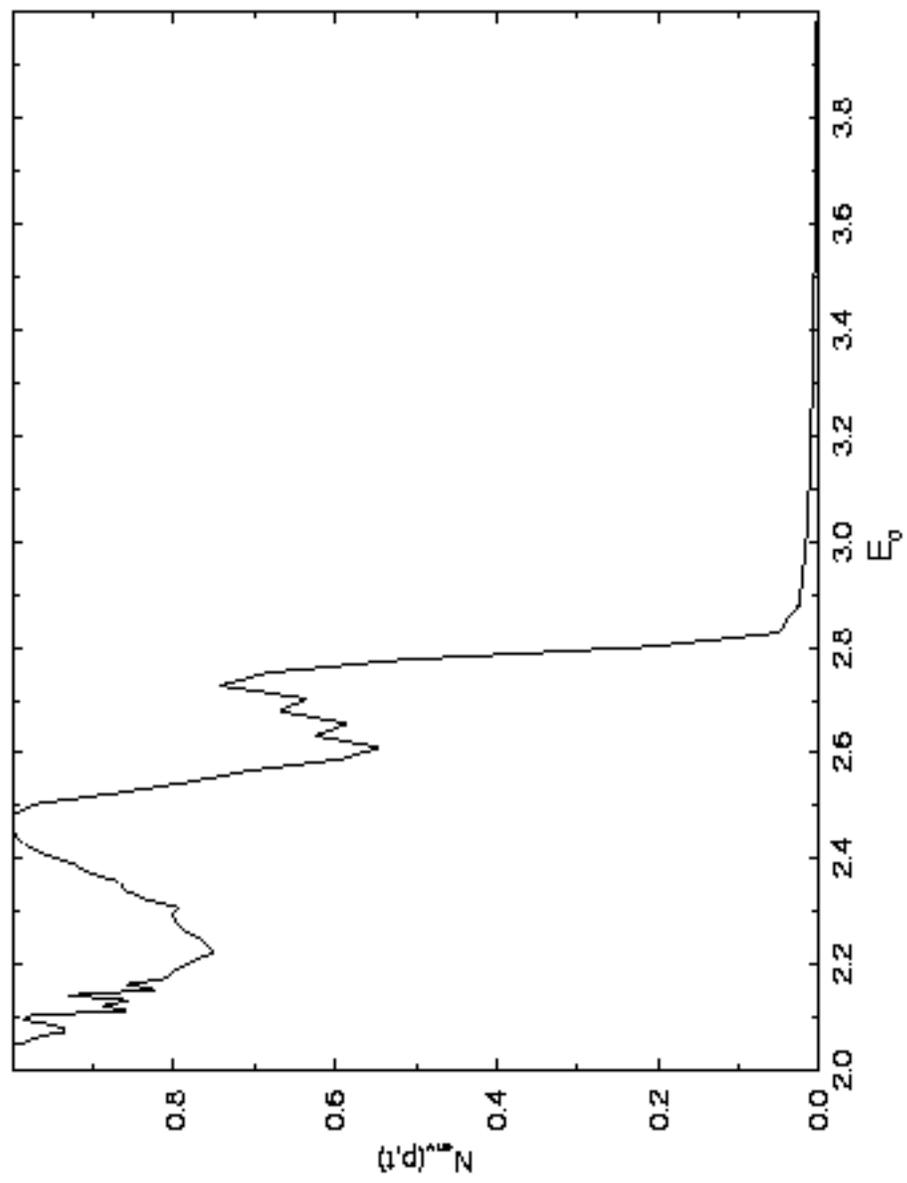
{\bf Fig. 13}: Maximal occupation number $\caln_{\rm env}$
 as a function of $E_0$ 
for the same parameters as in Fig. 11.
\\ \\
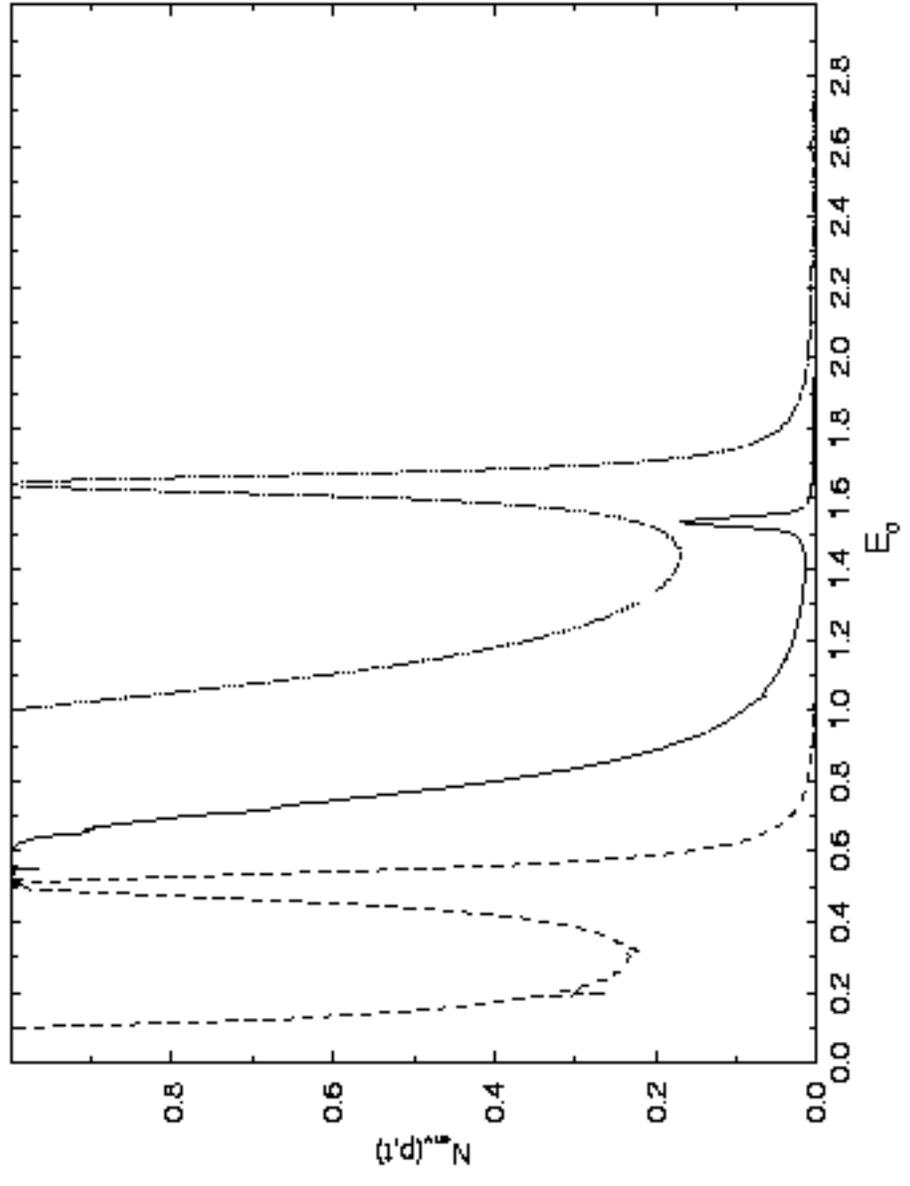
{\bf Fig. 14}: Maximal occupation number $\caln_{\rm env}$ for 
$m(t)=m_0\cos t$ with $m_0=0.1$ (dashed line),$m_0=0.5$ (solid line), 
and $m_0=1$ (dotted line), as a function of $E_0$.
\\ \\
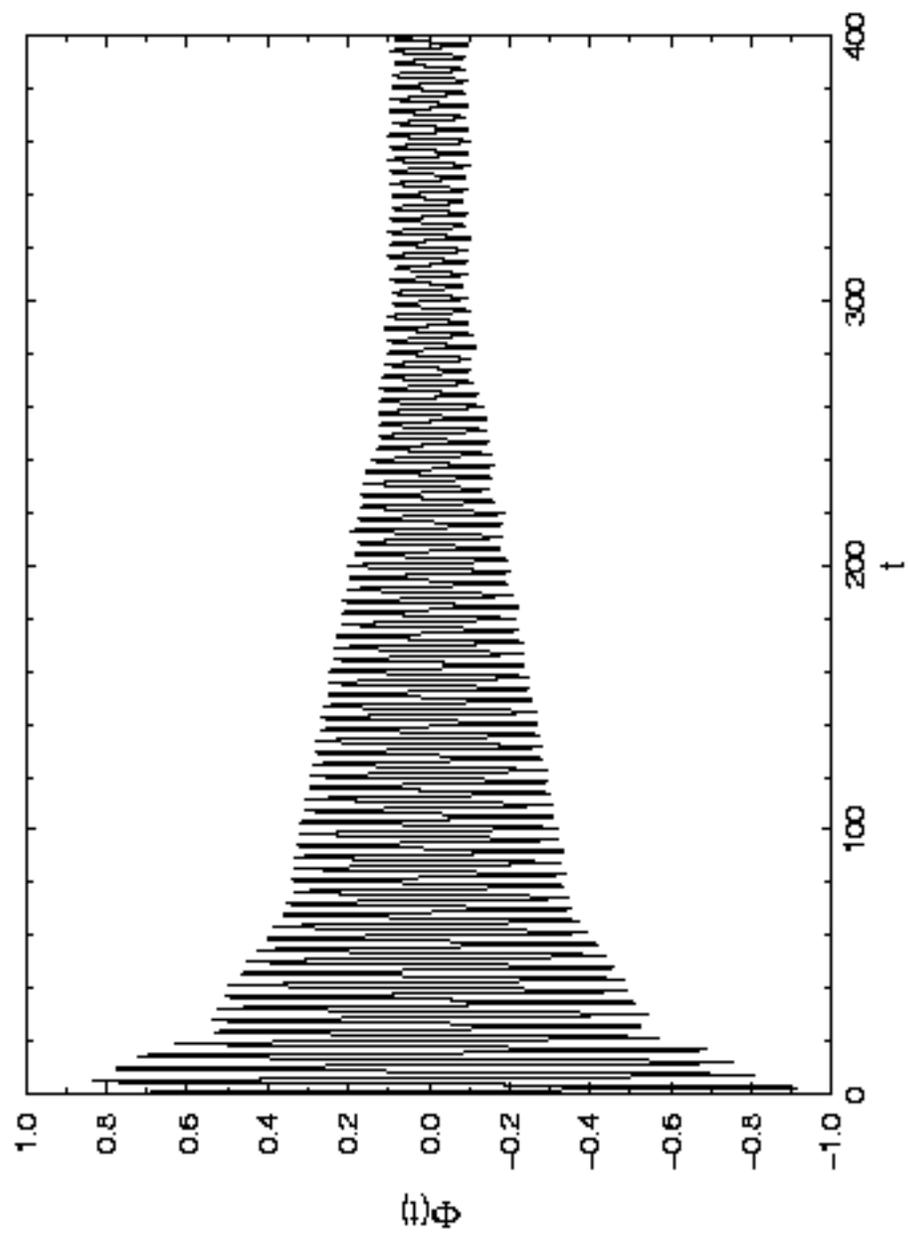
{\bf Fig. 15}: $\phi(t)$ including back-reaction of both fermionic 
and scalar
fluctuations for $M=1,g=2,\lambda=4$, and $\phi(0)=1,\dot\phi(0)=0$.
\\\\
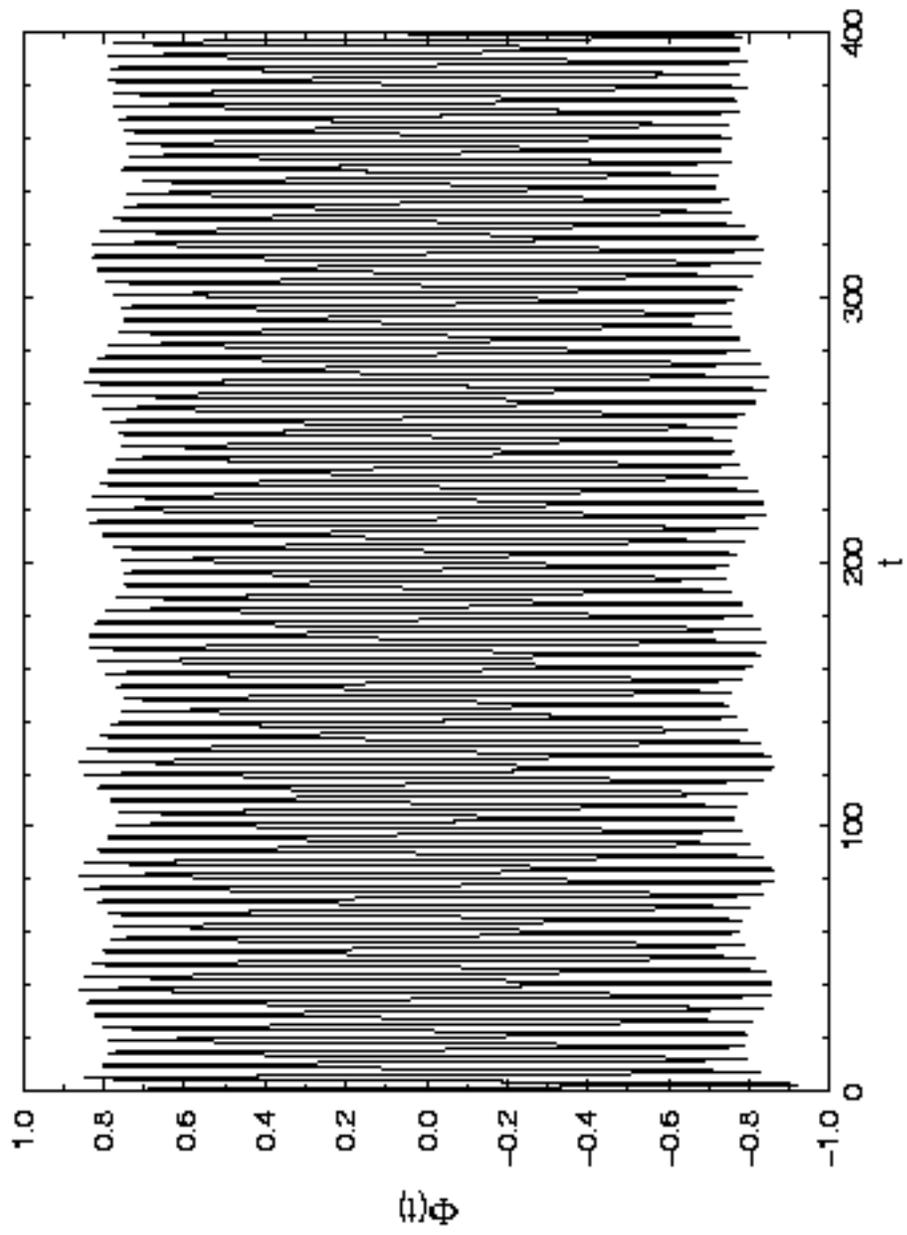
{\bf Fig. 16}: $\phi(t)$ in the absence of scalar fluctuations for the same 
parameters as in Fig 15.
\\ \\
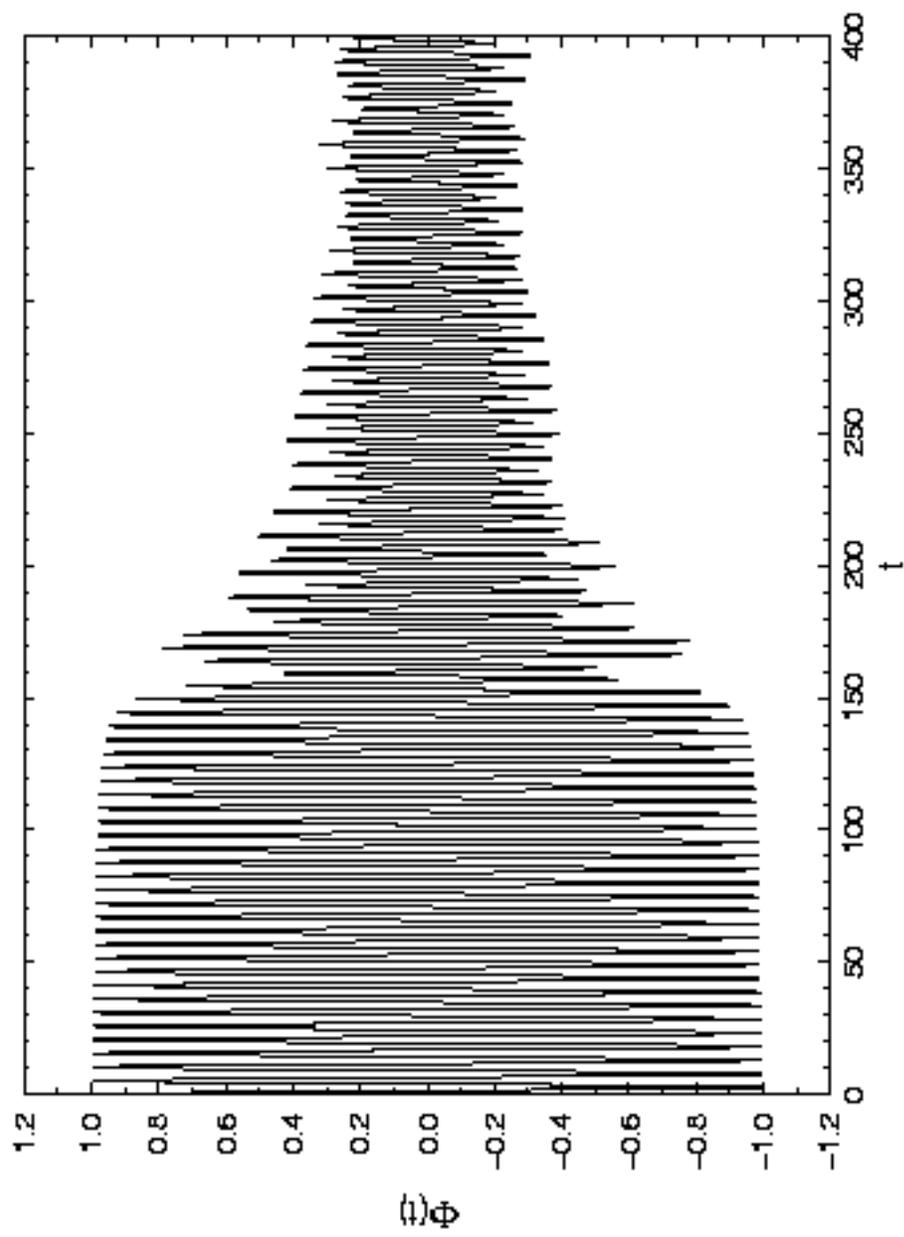
{\bf Fig. 17}: $\phi(t)$ in the absence of fermion fluctuations, 
for the same parameters as in Fig. 15.
\\ \\
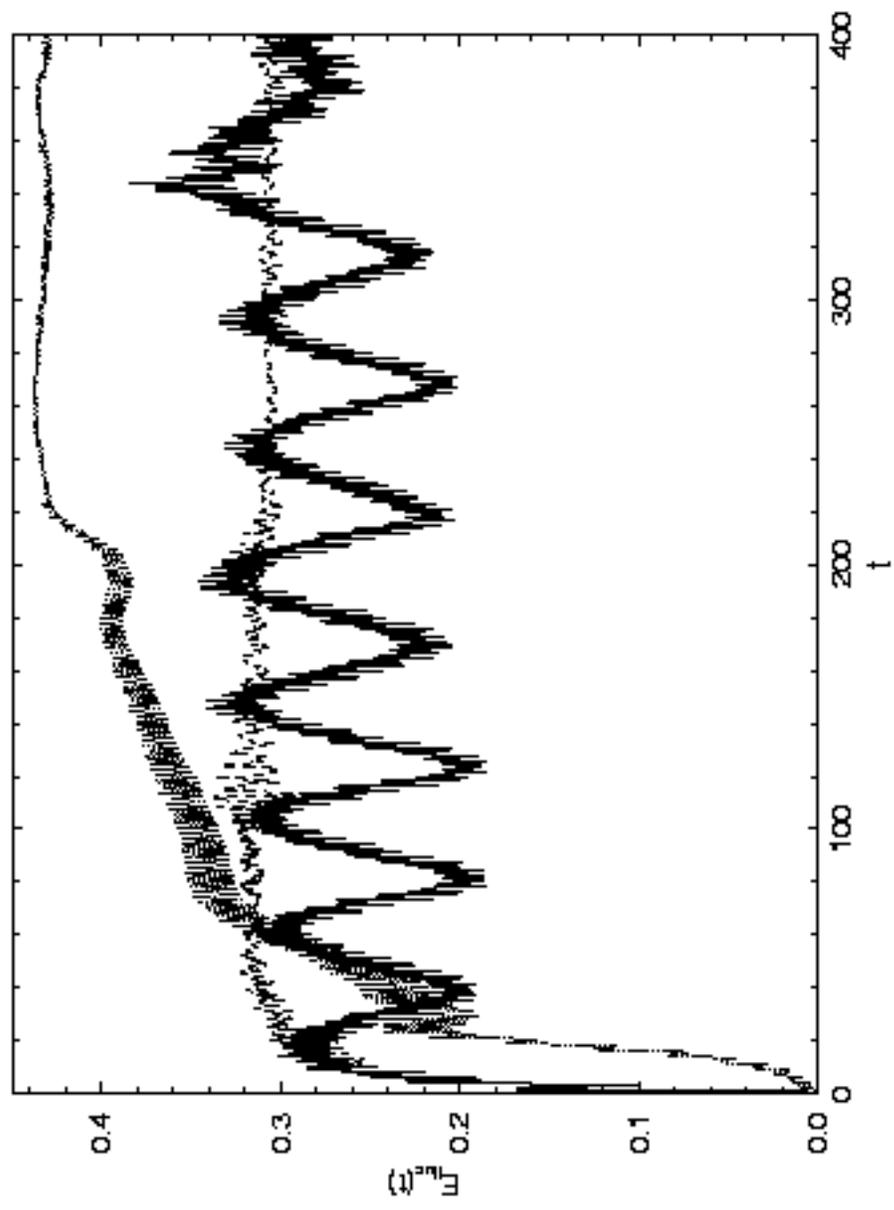
{\bf Fig. 18}: Fluctuation energies of 
fermions for evolution without scalar
fluctuations (solid line); fluctuation energies of
fermions (dashed line) and bosons (dotted line)
for combined evolution.


\begin{thebibliography}{10}

\bibitem{Ringwald:1987}
A. Ringwald, Z. Phys. C {\bf 34},  481  (1987); Ann. Phys. {\bf177}, 129
(1987); Phys. Rev. D {\bf 36}, 2598 (1987).

\bibitem{Calzetta:1987}
E. Calzetta and B.~L. Hu, Phys. Rev. D {\bf  35},  495  (1987);
{\em ibid} {\bf 37},  2878  (1988).

\bibitem{Kofman:1994}
L. Kofman, A. Linde, and A.~A. Starobinsky, Phys. Rev. Lett. {\bf 73},
  3195  (1994).

\bibitem{Boyanovsky:1994}
D. Boyanovsky, H.~J. de~Vega, and R. Holman, Phys. Rev. D {\bf  49},  2769
  (1994).

\bibitem{Shtanov:1995}
Y. Shtanov, J. Traschen, and R. Brandenberger, Phys. Rev. D {\bf 51},
  5438  (1995).

\bibitem{Khlebnikov:1996}
S.~Y. Khlebnikov and I.~I. Tkachev, Phys. Rev. Lett. {\bf 77},
219  (1996); Phys. Lett. {\bf B390},  80  (1997).

\bibitem{Son:1996}D. T. Son, Phys. Rev. D {\bf  54}, 3745 (1996).

\bibitem{Kofman:1996}
L. Kofman, A. Linde, and A.~A. Starobinsky Phys. Rev. Lett. {\bf 76},
  1011   (1996).

\bibitem{Boyanovsky:1997b}
D. Boyanovsky, D. Cormier, H.~J. de~Vega, and R. Holman, 
Phys. Rev. D {\bf 55}, 3373  (1997).

\bibitem{Boyanovsky:1997a}
D. Boyanovsky, D. Cormier, H. J. de Vega, R. Holman, 
A. Singh, and M. Srednicki, Phys. Rev. D {\bf 56}, 1939 (1997) .

\bibitem{Kofman:1997}
L. Kofman, A. Linde, and A.~A. Starobinsky, hep-ph/9704452 .

\bibitem{Greene:1997a}
P.~B. Greene, L. Kofman, A. Linde, and A.~A. Starobinsky,
Phys. Rev. D {\bf 56}, 6175 (1997).

\bibitem{Kaiser:1996}
D.~I. Kaiser, Phys. Rev. D {\bf 53},  1776  (1996);
{\em ibid.} {\bf 56}, 706 (1997).

\bibitem{Ramsey:1997}
S. A. Ramsey and B. L. Hu, Phys. Rev. D {\bf 56}, 678 (1997);
Erratum {\em ibid.} {\bf 57}, 3798 (1998).

\bibitem{Baacke:1997c}J. Baacke, K. Heitmann, and 
C. P\"atzold, Phys.  Rev. D {\bf 56}, 6556 (1997).

\bibitem{Boyanovsky:1998a}D. Boyanovsky, D. Cormier, H. J. de Vega,
R. Holman, and S. P. Kumar, {\em Out of equilibrium fields in inflationary
dynamics, density fluctuations}, hep-ph/9801453.

\bibitem{Kluger:1992}Y. Kluger, J. M. Eisenberg, B. Svetitsky,
F. Cooper, and E. Mottola, Phys. Rev. D {\bf 45}, 4659 (1992).

\bibitem{Boyanovsky:1995a}
D. Boyanovsky, H. J. de Vega, and R. Holman, Phys. Rev. D
{\bf 51},734 (1995).

\bibitem{Cooper:1994} F. Cooper, S. Habib, Y. Kluger,
E. Mottola, J. P. Paz, and P. Anderson, Phys. Rev. D {\bf 50},
2848 (1994).

\bibitem{Cooper:1995}F. Cooper, Y. Kluger, E. Mottola,
 and J. P. Paz, Phys. Rev. D {\bf 51}, 2377 (1995).

\bibitem{Cooper:1997}F. Cooper, S. Habib, Y. Kluger, and
E. Mottola, Phys. Rev. D {\bf55}, 6471 (1997).

\bibitem{Boyanovsky:1997d}
D. Boyanovsky, H. J. de Vega, R. Holman,
and S. Prem Kumar, Phys. Rev. D {\bf 56}, 1939 (1997);
{\em ibid.} {\bf 56}, 3929 (1997).

\bibitem{Baacke:1997b}
J. Baacke, K. Heitmann, and C. P\"atzold, 
Phys.  Rev. D {\bf 55}, 7815 (1997).

\bibitem{Boyanovsky:1998b}
D. Boyanovsky, H. J. de Vega, R. Holman, S. P. Kumar, and R. D. Pisarski,
 {\em Real time relaxation of condensates and kinetics
in hot scalar QED: Landau damping}, hep-ph/9802370.

\bibitem{Boyanovsky:1998c}
D. Boyanovsky, H. J. de Vega, R. Holman, S. P. Kumar, and R. D. Pisarski,
Phys. Rev. D {\bf 57}, 3653 (1998).

\bibitem{Zurek:1996} see, e.g., W. H. Zurek, 
Phys. Rept. {\bf 276} 177 (1996).

\bibitem{Schwinger:1961} J. Schwinger, J. Math. Phys. (N.Y.) {\bf 2},
407 (1961).
 
\bibitem{Keldysh:1964}
L. V. Keldysh, Zh. Eksp. Teor. Fiz. {\bf 47}, 1515 (1964);
Sov. Phys. JETP {\bf20}, 1018 (1965).

\bibitem{deVega:1997}H. de Vega and J. Salgado, Phys. Rev. D {\bf 56},
6524 (1997).

\bibitem{Boyanovsky:1997c}
 {\em Asymptotic dynamics in scalar field theory: anomalous relaxation},
D. Boyanovsky, C. Destri, H. J. de Vega, R. Holman, and
J. Salgado, hep-ph/9711384.

\bibitem{Boyanovsky:1995b}D. Boyanovsky, M. D'Attanasio, H.J. de Vega,
R. Holman, and D.-S. Lee, Phys. Rev. D {\bf 52}, 6805 (1995).

\bibitem{Ramsey:1998}
S. A. Ramsey, B. L. Hu, and A. M. Styrianopoulos,
Phys. Rev. D {\bf 57 }, 6003 (1998).

\bibitem{Baacke:1997a}J. Baacke, K. Heitmann, and 
C. P\"atzold, Phys. Rev. D{\bf 55},  2320  (1997).

\bibitem{Baacke:1998b}J. Baacke, K. Heitmann, and
C. P\"atzold, Phys. Rev. D {\bf 57}, 6406 (1998).

\bibitem{Stueckelberg:1951} E. C. G. Stueckelberg,
Phys. Rev. {\bf 81}, 130 (1951).

\bibitem{Bogoliubov:1980} N. N. Bogoliubov and D. V. Shirkov,
{\em Introduction to the Theory of Quantized Fields},
John Wiley and Sons, New York, 1980.

\bibitem{Baacke:1998a}J. Baacke, K. Heitmann, and
C. P\"atzold, Phys. Rev. D {\bf 57}, 6398 (1998).

\bibitem{Weinberg:1974}S. Weinberg, Phys. Rev. D {\bf 9}, 3357 (1974).

\bibitem{Callan:1970}C. G. Callan, S. Coleman, and R. Jackiw,
Ann. Phys. (N.Y.) {\bf59}, 42 (1970). 

\bibitem{Birrell:1982}
N.~D. Birrell and P.~C.~W. Davies, {\em Quantum fields in curved space}
(CUP,   Cambridge, 1982).

\bibitem{Barut:1987}A. O. Barut and I. H. Duru, Phys. Rev. D
{\bf 36}, 3705 (1987).

\bibitem{Bogoliubov:1983}N. N. Bogoliubov and D. V. Shirkov,
{\em Quantum Fields}, The Benjamin/Cummings Publ. Comp.,
Reading, Mass. 1983.


\end{thebibliography}
\end{document}